\documentclass[11pt, a4paper, logo, copyright, openany]{googledeepmind}

\usepackage[square,numbers]{natbib}
\usepackage{caption}

\title{ACAI for SBOs: AI Co-creation for Advertising and Inspiration for Small Business Owners}

\correspondingauthor{Nimisha Karnatak <nimisha.karnatak@some.ox.ac.uk>}


\reportnumber{001} 

\renewcommand{\today}{2025-03-09}

\author[1, 2]{Nimisha Karnatak}

\author[1]{Adrien Baranes}

\author[1]{Rob Marchant}

\author[1]{Tríona Butler}

\author[1]{Kristen Olson}

\affil[1]{Google DeepMind}
\affil[2]{University of Oxford}

\begin{abstract}
Small business owners (SBOs) often lack the resources and design experience needed to produce high-quality advertisements. To address this, we developed ACAI (AI Co-Creation for Advertising and Inspiration), an GenAI-powered multimodal advertisement creation tool, and conducted a user study with 16 SBOs in London to explore their perceptions of and interactions with ACAI in advertisement creation. Our findings reveal that structured inputs enhance user agency and control while improving AI outputs by facilitating better brand alignment, enhancing AI transparency, and offering scaffolding that assists novice designers, such as SBOs, in formulating prompts. We also found that ACAI’s multimodal interface bridges the design skill gap for SBOs with a clear advertisement vision, but who lack the design jargon necessary for effective prompting. Building on our findings, we propose three capabilities: contextual intelligence, adaptive interactions, and data management, with corresponding design recommendations to advance the co-creative attributes of AI-mediated design tools.
\end{abstract}

\begin{document}

\maketitle

\begin{figure}[h]
   \includegraphics[width=\linewidth]{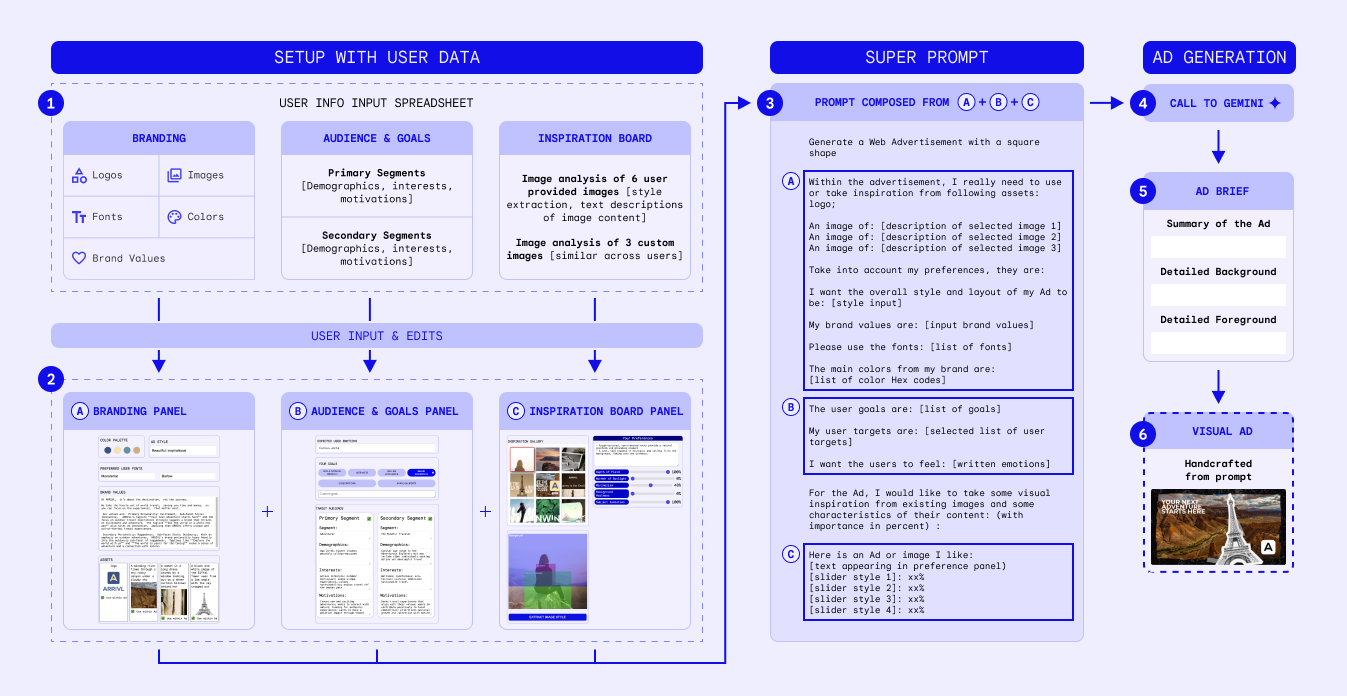}
   \caption{ACAI Prototype Design 1) User input submitted prior to the study 2) Interface design of ACAI incorporating user input and AI content into 3 panels 3) Backend "super prompt" concatenating user input 4) Call to Gemini to generate Ad Brief 5) Ad brief template 6) Visual ad designed by human designer according to Ad brief}
  
 \end{figure}

\section{Introduction}

Small business owners (SBOs) constitute a significant portion of the global economy, with 5.6 million in the UK alone representing 98\% of businesses \cite{forbesSmallBusiness}. Despite their economic importance, SBOs often face challenges in creating effective advertisements, which can make it difficult to showcase their brands and engage potential customers \cite{paradoxmarketingHatsWorn,undefinedWriteBusiness}. A recent survey of AI usage among UK small businesses revealed that 55\% believe AI could benefit their business\cite{fsbRedefiningIntelligence}, highlighting the potential of AI as a solution to their advertising challenges. Furthermore, several studies have looked at the application of Generative AI (GenAI) in advertising and marketing, underscoring its growing role in creative business processes \cite{10.1145/3491102.3517708,hartmann2024power}.

However, despite the promises, the current GenAI tools fall short in addressing the specific design needs of small business owners (SBOs). This gap manifests in two key tensions. First, most SBOs lack formal training or extensive experience in design, and therefore do not possess the specialized language or design terminology needed to craft effective prompts for AI-mediated co-creation. However, Generative AI tools are inherently dependent on prompts to generate effective outputs. This disconnect leaves SBOs, like other novice users, as noted in prior research \cite{johnnycantprompt_nonAIexperts, subramonyam2024bridging}, struggling to translate their creative ideas into prompts that AI systems can effectively interpret. Second, SBOs require advertisements that authentically reflect their unique brand identity in order to differentiate themselves in competitive markets. Current GenAI tools, however, frequently produce generic content~\cite{comedy_deepmind,Art_or_artiface_falsePromise} that could potentially fail to capture the brand-specific elements essential for distinctiveness. This lack of brand alignment risks diluting their message and reducing customer engagement, thereby diminishing the potential impact of AI-generated advertisements.

These misalignments reveal a gap, raising two central research questions that we address in this paper:
\begin{itemize}
    \item [1] How can GenAI tools be designed to effectively support novice designers, such as SBOs, in navigating the advertisement creation process?
    \item [2] How can GenAI tools ensure the generation of brand-aligned advertisements that authentically reflect the unique identity of small businesses?
\end{itemize}

To examine these questions, we developed a GenAI-based multimodal prototype called ACAI (AI Co-Creation for Advertising and Inspiration), specifically designed to assist novice users like SBOs in creating brand-aligned advertisements. ACAI provides a unified visual interface that centralizes branding assets. It incorporates three distinct panels—branding, audience and goals, and inspiration, each featuring predefined input fields for elements such as color palettes, fonts, brand values, target audience, and visual inspirations. This structured input mechanism is a core feature of ACAI, complementing its ability to interpret multimodal prompts. We conducted a user study involving 16 SBOs in London to evaluate ACAI’s effectiveness in supporting advertisement creation. 
Our findings demonstrated that ACAI supported novice designers through two key mechanisms: structured input scaffolding and multimodal prompting. The structured inputs guided users through the advertisement creation process by concatenating their inputs into "super prompts" for the AI, while multimodal prompting allowed users to extract text prompts from images. These features collectively reduced the cognitive load of writing a prompt, enhancing AI tool accessibility for novice designers. Additionally, ACAI's structured input interface facilitated the generation of brand-aligned advertisements by enabling user to incorporate specific brand elements, which enabled more accurate AI interpretation of brand identity. This approach resulted in outputs that authentically reflected each small business's unique characteristics. 

We advance the growing body of literature on AI-mediated creativity by focusing on novice users, specifically small business owners, in the context of generative AI tools for advertising in the following ways:

\begin{itemize}
    \item [1]  We conducted empirical research with small business owners, an understudied population in the HCI community, focusing on their interactions with generative AI for ad creation.
\item [2]  We introduce ACAI, a multimodal LLM powered ad co-creation tool, designed for novice designers specifically.
\item [3] We provide design recommendations to make AI a better co-creation tool, we recommend contextual intelligence, adaptive interfaces and data management.
\item [4] We illustrate how interface design can lower barriers for novice users and enhance user's brand alignment in GenAI tools, with implications for developing more inclusive AI systems that support novice users in complex creative tasks.
\end{itemize}

\section{Related Work}

\subsection{Theme 1: Emerging Practices in AI-mediated co-creation}
The discourse surrounding creativity within the HCI community has gained traction, particularly with the recent advancement in artificial intelligence. As creativity manifests as a complex, context dependent phenomenon, it precludes reduction to a monolithic, universally applicable definition. While traditional frameworks of creativity emphasize three core attributes- novelty, utility and surprise~\cite{Designing_Participatory_AI,traditionaldef_creativity,traditionaldef_creativity2}, recent empirical investigations on AI-mediated creativity have both challenged and expanded these human-centric constructs. 

 Consequently, researchers are exploring diverse perspectives on what constitutes AI-mediated creativity. \citet{genai_wild} found that creative professionals attributed value to Generative AI's (GenAI) large data processing and synthesis capabilities in their workflow, as a new form of creativity, even when the resulting outputs were not entirely original or lacked novelty. Similarly, ~\citet{Designing_Participatory_AI} revealed that creative professionals envision AI as a tool to expand their creative possibilities by rapidly generating diverse outcomes, thus enlarging their conceptual ``possibility space". Despite lacking a universal definition, HCI researchers have utilized and developed various methods to assess AI-mediated creativity. These include the Torrance Tests for Creative Writing (TTCW), an adaptation of the Torrance Tests of Creative Thinking (TTCT), for evaluating AI-generated short fiction~\cite{Art_or_artiface_falsePromise}; 
 and quantitative proxies such as remote clique scores~\cite{valueBenefitConcernOfGenAI}, Chamfer distance~\cite{valueBenefitConcernOfGenAI}, and semantic similarity analysis~\cite{CST_Homogenization_Analysis} for detecting homogenization in AI-assisted writing styles.

 Beyond quantitative assessments of creativity, HCI scholars have investigated the collaborative dynamics between humans and AI in creative tasks, a phenomenon conceptualized as `co-creativity.' Human AI co-creation definitions emphasize the responsiveness of an AI to user input ~\cite{identifyingethicalissuesai, 10.1007/978-3-030-78462-1_13} and the ability of humans and the AI to contribute to the creative output ~\cite{CaiCoCoCo}. To understand co-creativity, researchers have focused on a range of domains such as writing, comedy, art, and advertising to better understand  users' sense of creative freedom~\cite{comedy_deepmind}, the balance of control between user and AI~\cite{ghostwriterai}, development of shared understanding and trust~\cite{genai_wild}, and communication and feedback patterns~\cite{PromptCharm} that characterize these co-creative interactions. Within the context of marketing and advertising, Lyu et al.~\cite{Youtubers_Use_Of_GenAI} found that YouTube content creators leverage GenAI for a range of tasks, from content creation to managing affiliate campaigns. Complementing this, Kotturi et al.~\cite{VeneerofSimplicity_Entrepreneurs} explored GenAI's potential to enhance marketing strategies for resource constrained entrepreneurs.

 While these studies have advanced our understanding of GenAI's applications in marketing and entrepreneurship, a significant gap remains in HCI literature regarding the nuanced dynamics of AI-mediated advertising within small business ecosystems. Small business owners face unique challenges \cite{hbrSmallBusiness, berthon2008brand} related to resources, market positioning, and brand identity, which could critically shape their interaction with and perception of Generative AI in advertisement co-creation.

 Moreover, the tension between homogeneous AI-generated content ~\cite{ML_ArtistFolkTheories} and the need for distinctive brand identity in small business contexts offers a unique opportunity to investigate users' strategies for personalization and creative control. Our study aims to understand how small business owners perceive and interact with Generative AI in advertisement creation, as well as the perceived benefits and challenges of integrating AI into their workflow. By focusing on this understudied area, our work aims to provide insights that can inform the development of more effective and context-sensitive AI tools for advertising in small business settings.

\subsection{Theme 2: Designing for Novice Users in Human-AI co-creation}

Generative AI is rapidly transforming creative processes across diverse domains, leading to extensive investigation within HCI. A substantial body of research explores how expert users, such as writers~\cite{Art_or_artiface_falsePromise}, comedians~\cite{comedy_deepmind}, and film and theater professionals~\cite{dramatron_deepmind}, integrate and experience AI co-creativity tools in their workflow. This focus stems from researcher's interest in evaluating AI systems against expert benchmarks~\cite{Art_or_artiface_falsePromise} and exploring the ways in which these tools could reshape expert workflows and creative processes~\cite{opacity}.

Recent studies have also highlighted the challenges faced by novice AI users in effectively leveraging Generative AI tools. \citet{johnnycantprompt_nonAIexperts} revealed difficulties that novice AI users experience when crafting a prompt. Extending this line of inquiry,~\citet{subramonyam2024bridging} identified the `instruction gap,' where users struggle to convey intentions as explicit AI prompts, and emphasized that bridging this gap requires interface designs that seamlessly translate user intentions into effective AI prompts.

Recent studies have explored various methods to enhance LLM usability. For example, Liu et al.~\cite{10.1145/3613905.3650756} introduce Constraint Marker, a tool that lets users define constraints on output format, length, and style, improving readability and predictability. Similarly, Ma et al. present ExploreLLM\cite{10.1145/3613905.3651093}, which breaks down complex tasks into smaller subtasks using a structured UI, reducing cognitive overload for users in exploratory tasks.

However, a critical gap exists in our understanding of how individuals with varying levels of creative proficiency, particularly those without formal creative training or extensive practical experience interact with and adapt to AI systems for specialised creative tasks. 

Addressing this gap is essential for designing inclusive and adaptable AI interfaces that support users with varying levels of domain and AI expertise. We address this gap by investigating how novice domain users with diverse AI familiarity interact with Generative AI. We take small business owners, as a representative case study for interacting with ACAI, an LLM powered multimodal advertisement creation tool (as this group is uniquely positioned). ACAI´s structured input and multimodal interface simplifies prompt creation by leveraging user-provided assets and offering interface for multimodal and text-based prompt refinement. ACAI uses multimodal structured prompting to assist novice designers, specifically small business owners, in creating advertisements. Within ACAI, an inspiration board allows for the extraction of specific elements from an inspirational image. This is achieved by converting the element into text and including it in the prompt. This enables novice designers to include design elements they visually appreciate, even if they lack the technical vocabulary or expertise to describe or understand them.

\section{Study Design}

{To better understand the attitudes of using AI for creating advertisements, we conducted empirical research with small business owners in London.

\begin{figure*}[h]
   \includegraphics[width=\linewidth]{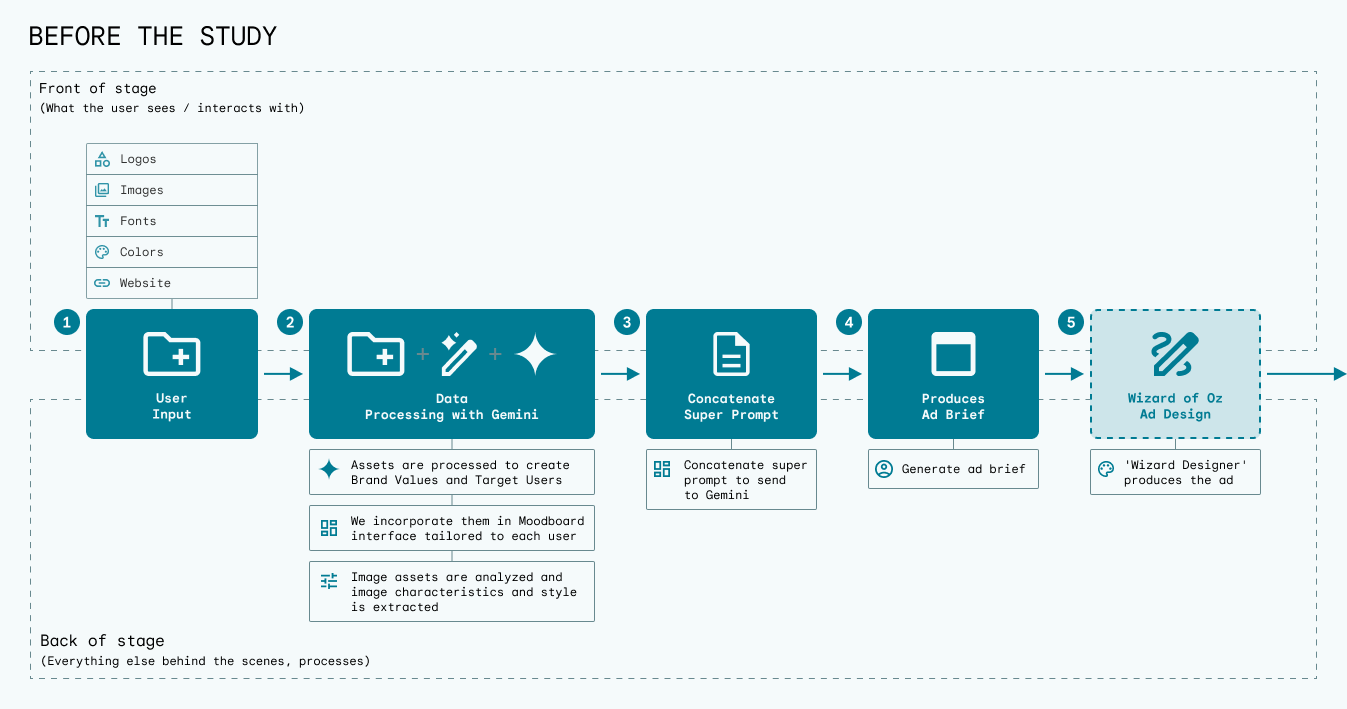}
   \caption{Before Study: 1) User Input 2) Data processing with Gemini 3) Concatenate Super Prompt 4) Ad Brief 5) Wizard of Oz design of Ad brief} 
 \end{figure*}

\subsection{\textbf{Before the Study}}
\textbf{Participants:}We conducted a user study with 16 small business owner participants in London. Participant business domains ranged from online retail to auto services and are detailed in Table 1. We had 11 male participants and 5 female participants whose ages ranged from 20 to over 61. Our participants were a mix of seasoned and new business owners with businesses from under 1 year to over 10 years ownership.
\textbf{Recruitment:} Participants were recruited from a database of individuals who were open to participating in research and were sent a screener form via email by a recruiter working with the team. Participants were eligible for participation if they were a small business owner \cite{forbesSmallBusiness} and 1) selected “brand awareness and perception” as one of the three most important aspects of their business they hoped to improve over the next three years and 2) consented to sharing branding assets to be used in an AI prototype for the study. All participants signed the organization's informed consent form prior to participation and the researcher reviewed the consent form with participants prior to the start of each session in person. The recruitment form specified that no branding assets would be used for AI model training. Each participant received a 105 pound gift card for their participation.\\
\textbf{User Input:} One week prior to the in-person session, participants uploaded a set of branding assets, including logo, colors, and reference images, to a Google Drive folder.\\
\textbf{Data Processing:} Given the variability in brand assets participants submitted, the study team organized assets in a CSV file to allow for consistent uploading into ACAI.\\
\textbf{Super Prompt:} Once the user input was uploaded, we used the structured input interface of ACAI to pre-populate the common participant goal of increasing brand awareness (identified as one of the top 3 business priorities) - in order to generate the Ad brief to be reviewed during the onboarding portion of the study.\\ 
\textbf{Onboarding Preparation:} For onboarding our participants to ACAI, we performed the following steps to develop the guided demo as detailed in Figure 3:
\begin{itemize}
    \item For the guided demo, an Ad brief containing a summary, detailed background, and detailed foreground description was generated from a call to Gemini.\\
    \vspace{-15pt}
   \item We used the Wizard of Oz method \cite{10.1145/169891.169968} to generate the visual advertisement for reviewing at the end of onboarding. A human designer created a visual advertisement corresponding to each participant's Ad brief for review during the study.\\ 
\end{itemize}
\vspace{-15pt}
In the following sections, we describe the interface of the ACAI in more detail, and the interaction of our participants with it during the study.
\subsection{ACAI Interface Design} 
{ACAI's design architecture is detailed in Figure 1. ACAI's structured input interface design was inspired by the constraints SBOs face in terms of design experience. With these constraints in mind, we structured the interface panels to include inputs that can influence high quality advertisements. Building upon this we hypothesized that having all the information in ACAI's structured interface would make it easier to include, exclude, or take inspiration from their brand assets-thus increasing the probability of brand-aligned advertisements.
We aggregated all participant input data (Brand Values, Color Palettes, Segments, Images) into a central spreadsheet exported to a CSV file. We then uploaded each participant file to the prototype interface prior to the study session. The ACAI interface consisted of 3 panels:}
\textbf{Branding Panel:} The business colors, fonts, and image assets, which consisted of the business logo and three images from the business. The ‘ad style’ was pre-populated by the research team and the ‘brand values’ were generated by Gemini 1.5 Pro Preview 0409 based on the participant data.\\
\textbf{Audience \& Goal Panel:} The panel consisted of three components. ‘Expected user emotions’ and ‘Your goals’ were pre-populated for the guided demo of the prototype. ‘Target Audience’ included primary and secondary segments informed by the business assets uploaded and expanded by Gemini.\\
\textbf{Inspiration Panel:} The final panel included an inspiration gallery that consisted of 3 existing business images and 3 inspirational images selected by participants, in addition to 3 images created by the team to highlight different visual styles e.g. cartoon, photo-realistic. For each image, we called Gemini before the studies to analyze each image to describe significant objects in the image and identify their respective coordinates, and call pre-written JavaScript functions to generate each rectangle with descriptions. We prompted Gemini to write the descriptions to a local file, loaded on demand during the study sessions.}\\
\textbf{Super Prompt:} The "super prompt" is a prompt that concatenates all of the user input into one, very detailed prompt to generate an Ad brief.\\
\textbf{Ad Brief Generation:} To generate the Ad brief, participants selected "Generate Ad brief" button. Behind the scenes, the super prompt of the user input from the three interface panels was sent to Gemini 1.5 Pro Preview 0409 to output an Ad brief. The Ad brief had 3 sections 1) Summary of the Ad 2) Detailed Background 3) Detailed Foreground. The Ad brief included these three sections to provide more specificity for the visualization of the advertisements by the design wizard. While GenAI can output visuals, we chose to focus participant feedback on the textual Ad brief and not use GenAI to generate the accompanying visual advertisement given the limitations of multimodal prompting at the time.\\
\textbf{Visual Ad:} Given technology constraints, ACAI was not able to generate a visual advertisement, instead all textual Ad briefs were given to a human designer to create a visual advertisement by reading the generated Ad brief and integrating the assets provided by participants as a proxy for multimodal prompting.\\

  \begin{figure*}[ht!]
  \includegraphics[width=\linewidth]{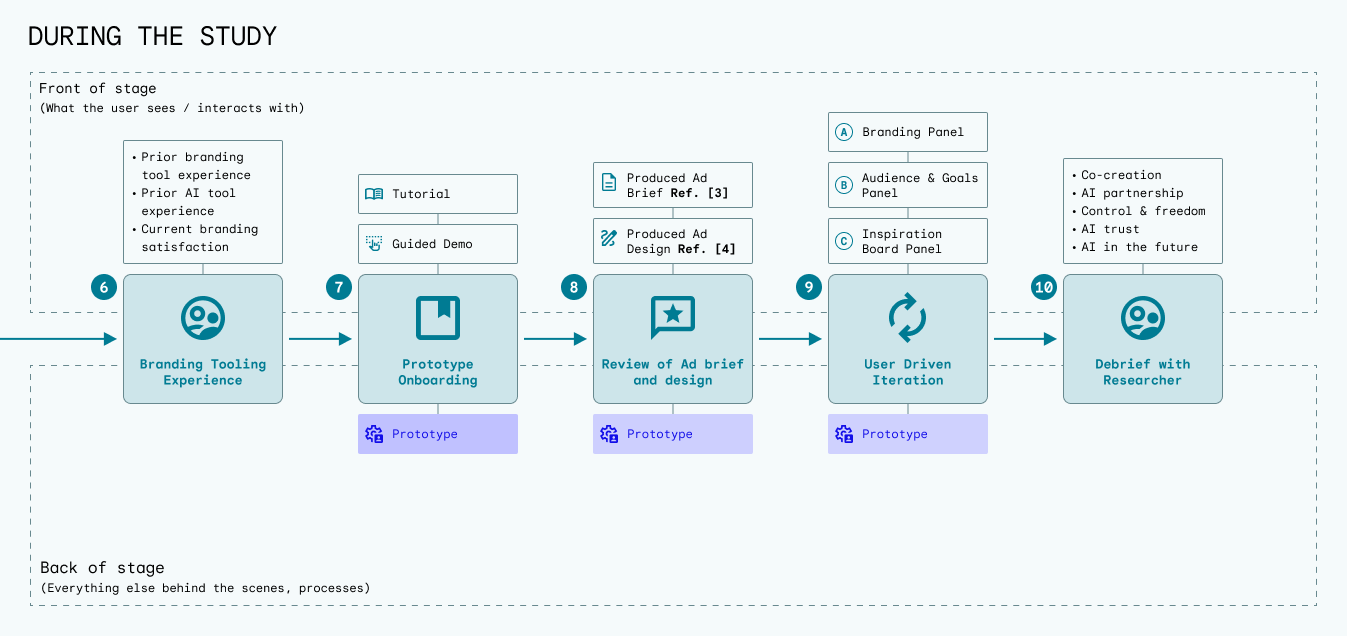}
  \caption{Study Flow Diagram: 6) Interview about using AI for business branding 7) Prototype onboarding 8) Review of Ad brief and Ad design 9) User driven iteration 10) Interview about overall experience and AI co-creation}

 \end{figure*}
\subsection{\textbf{During the Study}} 
\textbf{Branding tooling experience:} At the beginning of the study sessions, participants were asked about their experience with AI and non-AI tools for business branding.\\
\textbf{Scripted Tutorial:} Participants watched a four minute video demonstrating using ACAI to generate an advertisement for a demo business.\\
\textbf{Guided Demo:} Participants were guided through a demo with their business branding. The guided demo had each participant generate an advertisement with a `cartoon style'. While this style was not something that all participants would have chosen organically, we chose to apply the same demo experience for consistency between all participants.\\
\textbf{Review of Ad brief:} Participants reviewed the textual Ad brief generated based on the pre-selected options and then viewed the corresponding wizard of oz visual advertisement.\\
\textbf{Iterative Interaction:} Participants interacted with the prototype on their own and iterated on 1 - 3 Ad briefs. Participants did not see additional visual ad designs given the prototype was scoped to generate text Ad briefs only. \\
\textbf{Debrief with Researcher:} After generating and commenting on their final Ad brief, we interviewed participants about their overall experience and reflections on using AI for branding and co-creation.\\
\textbf{Analysis:} All study sessions were video recorded, with the exception of an error in P10's recording that resulted in only the first 20 minutes of the session being recorded. In addition to the recordings, the two researchers alternated facilitating sessions and the supporting researcher took notes. After all the sessions concluded, the two researchers met several times to analyze  and categorize participant responses into qualitative themes \cite{glaser2017discovery}.\\

\section{Results}
\begin{table*}[t]
    \centering
    \renewcommand{\arraystretch}{1.2} 
    \caption{Participant Overview: Demographic details of participants, their business domain, and current branding tool usage.}
    \label{tab:participant_overview}
    \resizebox{\linewidth}{!}{
    \begin{tabular}{|c|c|c|p{8cm}|} 
        \hline
        \textbf{Participant} & \textbf{Gender, Age Range} & \textbf{Business Domain} & \textbf{Branding Tool Usage} \\
        \hline
        P1  & M, 41-50  & Equipment Rental       & Adobe, Canva, Canva Pro, CapCut, Instagram, Mailchimp \\
        \hline
        P2  & M, 31-40  & Beverage Retail       & Instagram, Mailchimp \\
        \hline
        P3  & M, 31-40  & Auto Services         & Adobe, Instagram \\
        \hline
        P4  & M, 60+    & Professional Services & Adobe, Instagram \\
        \hline
        P5  & M, 51-60  & Healthcare Services   & Adobe, Instagram, Mailchimp \\
        \hline
        P6  & F, 51-60  & Gift Retail           & Adobe, Canva, ChatGPT, Instagram \\
        \hline
        P7  & M, 24-30  & Retail                & Adobe, Canva Pro, ChatGPT \\
        \hline
        P8  & F, 41-50  & Styling Consultant    & Adobe, CapCut, Instagram \\
        \hline
        P9  & M, 51-60  & Boat Services         & ChatGPT \\
        \hline
        P10 & F, 41-50  & Professional Services & Canva, Canva Pro, CapCut, ChatGPT, Instagram, TikTok \\
        \hline
        P11 & M, 18-23  & Online Retail        & Adobe, Canva, Canva Pro, CapCut, ChatGPT, Google Gemini, Instagram \\
        \hline
        P12 & F, 24-30  & Software Services    & Canva, Canva Pro, ChatGPT, Instagram \\
        \hline
        P13 & M, 31-40  & Experience Provider  & Adobe, Canva, Instagram, TikTok \\
        \hline
        P14 & F, 31-40  & Financial Services   & Adobe, Canva, ChatGPT, Instagram, Mailchimp \\
        \hline
        P15 & M, 41-50  & Food and Beverage    & Canva Pro, ChatGPT, Instagram, Mailchimp \\
        \hline
        P16 & M, 51-60  & Retail Goods         & Canva Pro, Instagram \\
        \hline
    \end{tabular}
    }
\end{table*}

\textbf{Current Branding Experience:} Our study revealed a range of tool preferences among participants for business branding. Among 16 participants, Instagram (87.5\%, n=14), Adobe Creative Cloud (62.5\%, n=10), Canva (43.75\%, n=7) and ChatGPT (43.75\%, n=7), emerged as the most frequently utilized tools. Six participants reported having previously created their own brand advertisements in addition to working with a designer while 4 created advertisements themselves and the remaining 6 outsourced to a designer or employee. To facilitate interaction with ACAI, we requested participant business brand assets, including a logo, preferred fonts and color, business website url, six images including previous advertisement or inspirational images. The 12 participants who had worked with a designer indicated that the materials they shared with us were comparable to what they shared with their designer previously. 
When asked to evaluate their satisfaction with their current branding, 11 participants reported being at least `somewhat satisfied' with their branding, while the remaining 5 were `somewhat dissatisfied'. 

Branding dissatisfaction was attributed to lack of time to develop desired assets and financial constraints faced by SBOs, which limited their ability to hire professional designers and to invest in multiple design iterations necessary for satisfactory results. P11 shared ``\textit{I don’t feel enough time gets put into it…we just don't have the manpower or the time or the budget essentially to put into this,"} and P16 described the difficulties of hiring a designer for one of his advertisements: ``\textit{she was very young and didn’t really understand the concept of the shop and what I was doing and to be honest with you, didn't really care.}"\\

\noindent
\textbf{Participant Experience With and Attitudes Towards AI:} We also explored if and how participants utilized AI tools within their business workflows. All 16 participants reported using AI to some extent in their businesses. While most were hesitant to trust AI for tasks like finance and calculations, citing concerns about its competence, they were more inclined to use AI for creative tasks such as business newsletters and social media captions. However, participants noted several challenges associated with using AI for creative purposes including: limited customization of AI outputs (``\textit{You can't edit the output retrospectively,}" - P7), generic results lacking brand authenticity (``\textit{It doesn't sound natural,}" - P10); ``\textit{(AI might present something that my business is not,}"- P16), plagiarism and SEO related risk 
(``\textit{never passed plagiarism tests,}" -P10), difficulty in prompting AI (``\textit{some text prompts that are kind of stale,}"-P1), negative perceptions of AI shaped by mainstream media and public opinion (``\textit{a Netflix program showed AI not recognizing people of color,}" - P5); ``\textit{(friends and colleagues consider using AI a cheat,}"- P8) and insufficient integration of new AI technologies with existing systems (``\textit{I want new technology to be integrated with the old}"- P16).

Despite these concerns, all participants demonstrated a positive outlook on the potential advancements in AI technologies to support their business, anticipating that future AI solutions would address current limitations. They expressed a willingness to explore AI's role in accelerating business operations, with the expectation that AI would offer solutions that are more cost-effective, faster, and easier to use to alleviate time, money, and expertise constraints.
\subsection{ Scaffolding Creativity in AI-mediated Design}
This section examined participants' experiences with ACAI's structured input interface, highlighting how it addressed challenges they encountered in previous AI interactions (as detailed above), while simultaneously providing scaffolding for a more user-centric co-creation.

\subsubsection{Structured Inputs for Enhanced Creative Control}
ACAI, featuring a three-panel interface (Fig. 1), enabled participants to systematically elicit and customize branding inputs, which were then synthesized into a ``super prompt'' to generate an AI-mediated advertisement. We found that this guided, structured inputs interface enhanced participants' sense of control and creative agency with P12 sharing, ``\textit{The change I see is in the structure of my involvement... I'm still in charge, I'm still leading the development of the creation... it's just giving me a better framework to follow.''} and P14 elaborating ``\textit{ I feel like I'm actually creating something more. Even though obviously I'm prompting the tool to do something, I feel like I’m having more inputs into the creative process."}

This suggests that for novice designers, a structured interface that scaffolds inputs to generate the final AI prompt could support the creative design process while preserving—and potentially enhancing—perceived user agency. It extends Vygotsky's concept of the Zone of Proximal Development~\cite{vygotsky1978mind} to AI-mediated design contexts, indicating that well-designed interfaces can serve as a form of cognitive scaffolding in design tasks as is supported in related literature ~\cite{Design_Collab_b/w_Designers_AI_LiteratureReview,  designpromptengineering, cocoDifussion}.

\subsubsection{Structured Input for Brand-Aligned AI Outputs} We also found that structured input, such as specific keywords, desired emotions, and brand values, provides a framework for the AI to understand the business brand identity. This results in AI outputs that are more likely to align with the brand's overall aesthetic and messaging as compared to a free-form text prompt with P2 summarizing, ``\textit{The fact that you get to have your asset and your inspiration, it does fill in that middle point of how the asset actually translates to what you're actually trying to put across."} P7 spoke to how the structured input addressed past experiences with generic AI outputs sharing, ``\textit{And I suppose if you put input all of this information it's hard for the outcome to not be unique. And I think that's what I like the most I suppose because if you do all that stuff then the thing is unique.} Similarly, P12 appreciated the clear categorization of structured input and noted that it addressed a common challenge in AI design tools: the difficulty of providing nuanced instructions. P12 shared, ``\textit{Usually, I guess you either have one or the other... when you tell AI include `this', it might not do it very well or it might make it too prominent. But having this distinction actually between assets and inspiration board... I can really detail whether I specifically need you to include `this' in it, or I want you to be inspired by `this'."} Thus, structured inputs allowed our participants to communicate their design intentions more effectively leading to more alignment with their brand vision and AI output.

\subsubsection{Structured Input for Improved Transparency}

Our findings demonstrated that the structured nature of the interface, with its defined input fields, allowed participants to clearly observe how modifications in individual input fields directly influenced the final output. This increased transparency contributed to heightened perceptions of user agency and control. As participant P7 observed, ``\textit{It is very sensitive, it is really responsive. I know there's that (pointing to an element in output) from the edits I made."} 
However, participants also expressed a desire for increased granular control and a deeper understanding of the AI's decision-making process. For example, P12 suggested implementing input weighting mechanisms to modulate the influence of individual inputs on the final output. Additionally, all participants strongly advocated for real-time output updates in response to input adjustments, emphasizing the need for an interface that provides immediate visual feedback to help users understand how their actions shape the design outcome. 

In summary, the idea of ``structured co-creation'' encapsulates the balance between user control and AI assistance, facilitated by the interface. The interface scaffolds the process through structured inputs while the user remains firmly in the driver's seat, leading the creative process and making the final decisions.

\subsection{Multifaceted Co-creation with ACAI}
In this section, we discuss participant´s understanding of co-creation with AI, how ACAI mediates co-creation through multimodal prompting, participant´s experience of co-creation with ACAI, and the consequential increase in user agency in co-creation.
 \subsubsection{Unpacking User Perceptions of AI Co-Creation} Our study investigated user perceptions of co-creation with AI, specifically what co-creation meant to them and whether they felt ACAI facilitated co-creation with AI. An analysis of participant interviews revealed diverse perspectives, for example, P4 and P15 viewed co-creation as a catalyst for creative exploration, allowing them to explore novel ideas.
 P8 and P16 emphasized user agency, perceiving co-creation as a means to guide the design process and maintain control over the final output. P12 and P7, however, framed co-creation as a collaborative process, highlighting the crucial role of their input and guidance in shaping the final creative outcome with the AI. After interacting with ACAI, P12 shared, ``\textit{While tools like ChatGPT often feel limited and rigid, this prototype feels different—it adapts to my inputs, responds to my changes, and evolves with my feedback. To me, that really feels like co-creation.}"
 
 \subsubsection{Multimodal Prompting in AI-Mediated Co-Creation for Novice Designers} Despite a strong intuitive understanding of their business brand identity and desired advertisement designs, our participants, as novice designers, often struggled to bridge the gap between their creative vision and the precise technical design jargon required by AI systems for effective co-creation. To address this challenge, we introduced an Inspiration Board panel in ACAI, facilitating multimodal prompting. This feature allowed participants to upload an inspirational image and extract AI-generated technical design descriptors from selected visual elements within the image. We found that this functionality enhanced our participants' ability to engage with the AI system in a more meaningful and technically precise manner with 10 out of 16 participants highlighting the inspiration capability as the most exciting and helpful feature of ACAI. For instance, P15 appreciated the Inspiration Board's utility, describing it as ``\textit{very cool,}'' while P4 emphasized its practicality: ``\textit{I mean, writing prompts is difficult, right? So yeah, the fact that you were able to extract the prompts from the image is a big help.}''

 We believe this approach embodies co-creation in its true sense: users contribute their contextual understanding of brand identity, while the AI provides technical design descriptions of visual elements to user. This allows the user to refine these descriptions, which the AI then aggregates into a super-prompt to generate the final design output. Thus, both work in tandem, leveraging their respective strengths to achieve optimal design outcomes.

\subsubsection{Source of Inspiration and Rapid iteration}
Participants appreciated ACAI as a platform to create advertisements in addition to giving them the ability to iterate quickly upon their ideas. For example, P9 appreciated how ACAI tool could ``do the convergent thinking" while they focused on ``divergent thinking" and P15 saw ACAI as a ``\textit{brilliant idea generator."}

Participants identified a range of potential applications for ACAI beyond creating advertisements.
P14 appreciated the feeling of being in the "driver's seat" and noted that he ``\textit{would be able to create many more campaigns and hundreds of different versions of the same image or campaign...and then A/B test them to see which one performs better."}

P1 further envisioned using ACAI to pre-program content and campaigns tailored to specific dates such as Pride Month and Halloween. Beyond streamlining marketing tasks, participants also identified ACAI's potential to enhance other aspects of their business's online presence, with P8 planning to use it for crafting her website's "About Us" section, impressed by its on-brand descriptions. Our study also revealed that participants, when interacting with ACAI's structured input interface, were able to generate more comprehensive marketing strategies. Through iterative refinement of brand values, target segments, and brand-aligned visuals, participants demonstrated a broadening of their marketing perspectives. For instance, P12 shared, ``\textit{Most of my ads focused on just selling a product, but now I try to draw customers to my Etsy page before selling to them."} This suggests that ACAI nudges SBOs toward adopting a more holistic marketing approach, encouraging strategies beyond direct sales, such as fostering customer engagement. Thus, by recognizing ACAI's potential beyond creating advertisements, participants engaged in a collaborative process of adapting and shaping the technology to better suit their specific needs and goals.

\subsubsection{ACAI Supporting Human Expertise}

While all participants found ACAI easy to use, regardless of prior AI experience, those who had experience creating their own advertisements exhibited greater confidence modifying inputs in ACAI. For instance, P7, who is currently pursuing a business degree while building his own brand, remarked, ``\textit{Business branding is crystallized in my head; about what I want the brand to be and how I want to achieve it."} We observed that this clarity enabled P7 to provide more focused input and to effectively leverage ACAI’s customization options to reach desired output. 

We also investigated whether participants perceived ACAI as a replacement for or a complement to professional designers, while being mindful that the majority of our participants did not have regular access to a designer. The majority (9 out of 16) regarded it as a complementary tool. P2, for example, envisions using ACAI ``\textit{with the designer.}" This aligns with P7's idea of using ACAI to handle the ``\textit{initial briefing stage}'' of ad, to make his ideas clearer before communicating with a designer.
 
P14 speculated that ACAI's complementarity could influence future handling of design tasks, stating, ``\textit{with the prototype... I can do a lot of these kind of basic tasks myself .... I only need to go to them (designer) when I need really that kind of high-quality work and 80\% of what I do I can do it myself.}" Our findings suggest that ACAI could redefine the design process for small businesses by streamlining task distribution, enabling SBOs to manage routine tasks independently while relying on designers for complex creative challenges, and fostering better communication between SBOs and designers by augmenting their ability to think strategically and articulate design requirements effectively. This shift indicates a transition in small business owners' roles from passive clients to active participants in the design process.

\subsection{Ambiguities Around AI Partnership and Data Ownership}
Participants agreed that ACAI contributed to co-creation in several ways as detailed in the previous section. However, due to rapid advancement in these emerging technologies and given the limited bandwidth of SBOs, we found that participants were more concerned about the quality of the final output rather than understanding the inner workings of AI systems. This led to ambiguities regarding AI´s agency, leading to conflicting views about its status as a partner or a tool and safe data handling in the co-creative process.

\subsubsection{Is AI a Collaborative Partner?}
We explored participants' perceptions of AI as a collaborative partner in the design process. We found that nine of our participants considered AI as a partner in the context of ACAI, citing reasons such as ACAI's responsiveness, input interpretation, and ability to generate brand-aligned output. However, seven participants disagreed, expressing hesitation or outright resistance to the term ``partner.''

For example, P9 remarked, ``\textit{It is not a person!}'' while P14 compared AI to a pen, suggesting that calling it a collaborator attributes too much agency to it. These participants primarily viewed AI as a ``\textit{sophisticated tool}'', with P4 explicitly describing it as such.

When asked what would make AI a collaborative partner, participants highlighted three key factors beyond competency and transparency. First, AI should have the ability to retain and recall past interactions to understand their specific business nuances, eliminating the need to repeatedly convey established information. 
As P11 noted, ``\textit{ My designer already got that (business) context. I don't need to go through it every time with him.}'' 

Second, participants (P1, P11, P14) emphasized the importance of long-standing relationships with their designers, who understand their preferences and know how to translate them into tangible outcomes. P1, who has worked with the same designer for 15 years, explained, ``\textit{We know intimately how things work. I can go to him with a feeling about something, and we both know the brand guidelines in our heads.}''

Finally, participants desired AI to refine and improve the outputs by offering design feedback, like an experienced designer or marketing expert, to enhance the sense of collaborative co-creation. P12, for example, shared, ``\textit{I would find it amazing if I could finish kind of an initial draft ... and I could ask [the AI] to almost review or give me your opinion on what you think about the ad and what you think should be improved.}"

Regardless, even participants who primarily viewed AI as a tool expressed willingness to incorporate ACAI into their workflows, appreciating its ability to generate brand-aligned ads quickly, with cost being the main factor they would assess before integration.

\subsubsection{AI Data Concerns}

Our participants demonstrated uncertainty regarding inputting data into AI tools and the status of copyright and AI-generated content. P14 summarized the ambiguity of inputting data, particularly proprietary business data into an AI tool sharing, ``\textit{I mean, well one big thing is obviously what happens to your data. That's the main thing with any AI tool. So for us when I'm doing things for general marketing I think it's okay to send things, stuff that is already in the public domain.}'' While P4 expressed a valid concern about the risk of copyright infringement when using ``\textit{stock photos}'', P5 supposed that AI could circumvent these concerns, stating, ``\textit{If AI can create any image, then we won't have to worry about copyright.}'' This assumption highlights a broader uncertainty between the authorship of prompts and the ownership of AI-generated content— a theme noted in prior literature~\cite{AI_art_and_impact, 10.1145/3614407.3643696, isITai_orMe, designfictionstudy}. P10 acknowledged this lack of clarity and  reflected, ``\textit{There's a lot of people who don't know what they don't know, myself included. I feel that education is the key, like how to use it (AI) in an ethical, good way that still represents your business well.}'"

Related, without clarification on how the AI system is leveraging data input into the model by all users, some participants, like P16, expressed concerns about whether outputs from ACAI would result in similar outputs for everyone and make it harder to stand out. However, participants also spoke to instances where competitors had copied branding and advertising styles with P1 sharing ``\textit{We found competitors just jumped on any advertising we put out there, they just jumped on and started sucking (branding style) it off and everything."}

\section{Discussion}

Our study revealed that small business owners envision AI in design processes not as a replacement for designers, but as an augmentative tool.  Based on these insights, we propose three capabilities to strengthen AI's role as a complementary tool within small business workflows: contextualized business memory, adaptive interface, and data management.

\subsection{Contextualised Intelligence}

\begin{figure*}[t!]
   \includegraphics[width=\linewidth]{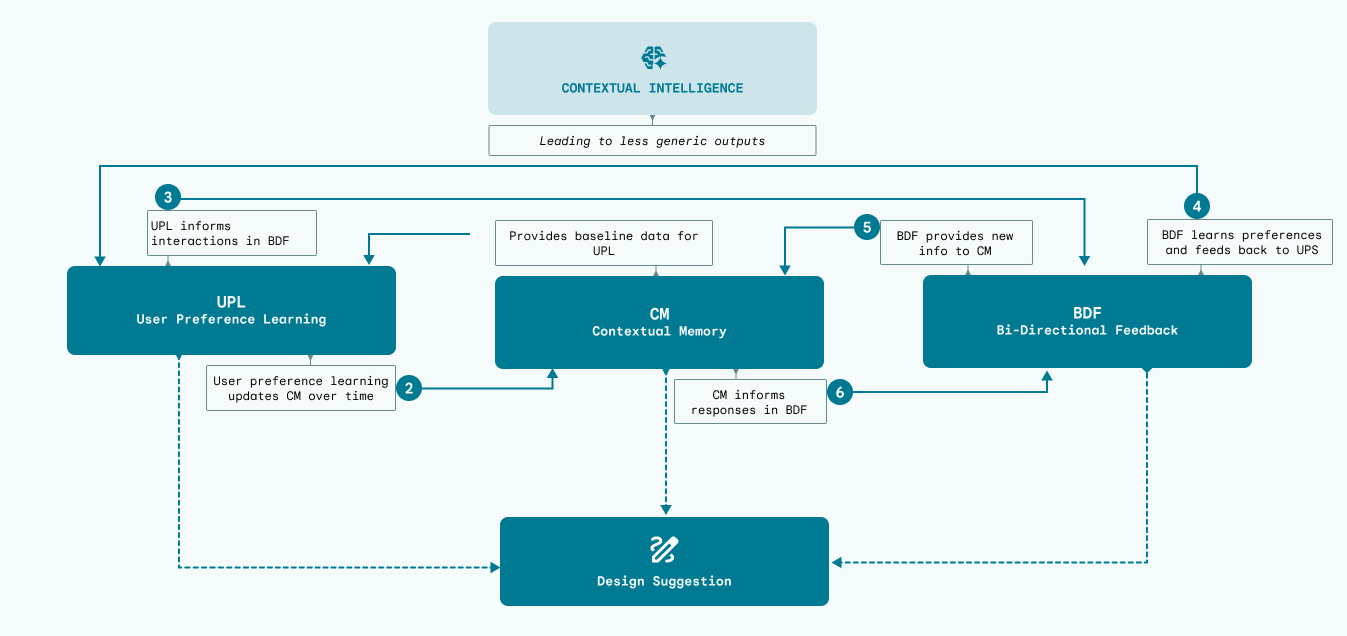}
   \caption{Contextual Intelligence - Design recommendations for less generic output 1) Provide baseline data for user preference learning (UPL) 2) User Preference Learning updates Contextual Memory (CM) over time 3) UPL inform interactions in Bi-Directional Feedback (BDF) 4) BDF learns preferences and feeds back to UPL 5) BDF provides new information to CM 6) CM informs responses in BDF. This operates as a dynamic, cyclical process where each component informs and is informed by the others, creating an adaptive and responsive AI-mediated design process within the framework of Contextual Intelligence.} 
   \label{fig4}
 \end{figure*}

We found that participants who had worked with designers previously emphasized that AI should recall and leverage past interactions, understand their specific business preferences, and provide feedback on design outputs to be a more effective tool. In response to these needs, we propose a \textit{Contextual Intelligence} framework for AI-mediated design systems. While the term ``Contextual Intelligence" has its origins in leadership studies~\cite{contextualintelligence}, we adapt this concept for application in AI-assisted design. Our framework operationalizes \textit{Contextual Intelligence (CI)}  through three key components: Contextual Business Memory, User Preference Learning, and a Bi-directional Feedback Mechanism. The interplay of these components is illustrated in Figure \ref{fig4}.

\subsubsection{Contextualised Business Memory}
To enhance the AI's understanding of business-specific needs, we recommend implementing a contextual memory system that extends beyond storing basic brand assets. This system should retain and recall past design decisions on layout, styling, and content placement; SBO feedback history and reactions; specific advertisement goals; and longitudinal advertisement agendas. 
Here, we are proposing actionable design recommendations for AI systems that cater specifically to SBOs. We recommend a memory system should include:

\begin{itemize}

\item \textit{Long-Term Memory for Brand Consistency:} stores high-level summaries of brand values, target audiences, and visual styles. This long-term memory should act as a cognitive scaffold, aiding SBOs in maintaining a consistent and coherent brand identity across multiple projects.
\item \textit{Cross-Session Design Recall:} stores specific design choices made across different sessions within a project or across related projects. This includes layout, color, fonts etc.

In essence, we foresee long-term memory storing high-level brand elements, ensuring overall consistency, and maintaining a timeless brand identity. Cross-session design recall, on the other hand, retains project-specific design decisions, facilitates continuity across related sessions, and supports rapid design iterations within the current context.

\item \textit{Temporal Awareness:} ability to evaluate design elements based on their relevance over time. Previous studies have focused on using temporal encoding \cite{zhong2024memorybank,myagentunderstandmebetter} to assess the historical relevance of assets and decay functions to gradually reduce the influence of outdated design elements \cite{zhong2024memorybank,myagentunderstandmebetter}. 
 
We recommend modifying the decay function to incorporate user preferences, preventing the system from discarding older design elements solely based on their age \cite{zhong2024memorybank}, thus mirroring how human memory prioritizes information based on personal significance.
\item \textit{Real-Time Feedback Storage:} capture and store real-time feedback from both the user and, if applicable, the AI. This will add to the baseline data for user preference learning (See Figure 5).

\end{itemize}

\subsubsection{Learning User Preferences}
We recommend user preferences to be stored in Contextual Memory, allowing users to consistently access their preferences across sessions. Based on feedback from the user, these preferences are updated with each interaction, which also refines the temporal awareness of the Contextual Memory. User preference learning plays a crucial role in personalizing AI feedback, ultimately shaping the design process \cite{10.1145/3477495.3531873,azar2023generaltheoreticalparadigmunderstand,10.1145/3604915.3608845}.
Recent studies indicate a shift from traditional, opaque representations of user preferences, such as those used in collaborative filtering, Reinforcement Learning from Human Feedback (RLHF),Direct Preference Optimization (DPO) towards more transparent methods like scrutable preference learning \cite{10.1145/3477495.3531873}. Scrutable preference learning allows users to inspect and modify the AI's understanding of their preferences \cite{10.1145/3477495.3531873}. This approach is especially valuable for small business owners and novice designers, as scrutable systems present learned preferences in easily understandable formats, such as natural language summaries or visual representations \cite{10.1145/3477495.3531873}. For example, SBOs could see a dashboard displaying their input style preferences, favorite layout types, or commonly used design elements. Additionally, scrutable systems provide explicit feedback based on observed user actions \cite{10.1145/3477495.3531873,10.1145/3604915.3608845}. Users can confirm or correct these interpretations, creating a feedback loop that enhances the AI's understanding \cite{10.1145/3477495.3531873,10.1145/3604915.3608845}. This transparency allows users to refine the AI’s understanding, leading to improved personalization over time.

\subsubsection{Bi-Directional Feedback}
Our findings indicate that SBOs seek bi-directional feedback with AI systems during the advertisement creation process. We propose moving beyond feedback alone towards a more integrated co-creation process and present a framework based on the three stages of generative AI collaboration — definition, generation, and reflection — as outlined in prior literature~\cite{cocoDifussion}. Our approach highlights the critical role of interface design in facilitating this collaboration \cite{fromPrompt_toCollab}.
We build upon the growing body of literature examining the role of interface design for user-centric AI tools \cite{10.1145/3643834.3661588,10.1145/3613905.3651093, PromptCharm, 10.1145/3614419.3644000}.
\begin{itemize}

\item During the definition stage, we envision AI offering brand-aligned suggestions to users by leveraging contextual memory. This interaction could be facilitated by an idea generation panel—a collapsible, multimodal input interface, primarily used at the outset of the design process or when the user requires broader, strategic input. In the future, we see AI leveraging contextual memory to pre-generate ads, allowing users to access and refine them immediately upon logging in to address SBO time constraints.

\item Our findings demonstrate that ACAI's interface provided users with creative control, revealing opportunities for future systems to offer multimodal prompts and scaffolding tailored to individual expertise and preferences. 
 
We also found that well-designed interfaces can facilitate better co-creation with AI, without compromising creative control. Looking forward, we envision AI-mediated design tools leveraging contextual memory to provide element-specific, real-time assistance during creative process. For instance, users can select individual elements, receive tailored AI feedback through a minimally intrusive UI component and integrate these insights, all while maintaining control over their creative process.

\item Our participants wanted AI to provide feedback on the output, similar to a designer, to make the process feel truly co-creative. In the reflection stage, we believe AI should offer feedback grounded in design principles.
While we recognize that users may often expect one-off results—advertisements that are immediately satisfactory without further feedback, we understand that even when following design best practices, individual preferences can vary. Offering insights into how specific elements or their placement are based on design principles could empower SBOs to modify the design to better suit their needs.

\item Users can also ask specific design-related questions during the reflection process, with AI supporting its responses referencing sources found online. Throughout the session, any AI suggestions accepted by the user will be integrated into the system’s memory, enhancing it and enabling the AI to learn user preferences. Future research could explore more nuanced methods of capturing feedback, such as when and if users prioritize feedback that helps them achieve immediate goals (effectiveness) or feedback that introduces new perspectives (novelty) could lead to more intelligent and adaptive AI systems.

\end{itemize}

\subsection{Adaptive Interface}
\subsubsection{Fluid Control}

Our study uncovered a nuanced relationship between user input and perceived control in AI-assisted ad creation. While participants generally equated a higher number of inputs with increased control, this relationship emerged as non-linear and multilayered. For example, one participant highlighted the importance of prioritizing inputs to enhance efficiency and suggested assigning weights to inputs to improve transparency. These suggests that the optimal number of input and their level of detail required for optimal control may vary based on specific use cases and contextual demands. While existing literature underscores the importance of user control in AI design \cite{If_the_Machine_Is_As_Good_As_Me_Then_What_Use_Am_I?}, translating this into practical solutions that can adapt to users' evolving input preferences and needs throughout the design process remains a challenge.

To address this, we propose the concept of \textit{fluid control}, where the AI dynamically adjusts its assistance in real time based on insights from past research on user behavior, task complexity, and evolving needs when using AI-mediated tools. Prior literature suggests that novice users prefer providing less input to achieve results with minimal effort~\cite{lumimood}. Conversely, expert users tend to prefer more input, allowing them to exert greater control and refine their designs~\cite{lumimood}. Task complexity influenced this relationship as well; intricate or technically demanding tasks required more user input for control, whereas simpler or exploratory tasks benefited from less input, allowing the AI to contribute more creatively~\cite{NonLinearCollaboration}. User goals further shape this balance, with those prioritizing efficiency, like our participants, opting for minimal input without compromising output quality, while those focused on creative exploration provide more input to retain control over the process~\cite{postprocessing}. Throughout the design process, input varied—less during initial ideation to leverage AI assistance~\cite{tooLatetobecreative,Divergent_Thinking}, more during refinement for detailed control~\cite{tooLatetobecreative}.  
Trust in AI significantly affects user input. Users with higher trust rely more on automation with less input, while lower trust leads to a preference for manual control~\cite{trust}. These factors, as established in prior studies, make the relationship between user input and control multi-dimensional, highlighting the need for a fluid, adaptive approach that accommodates the dynamic nature of design tasks and user preferences.

We propose developing AI tools that leverage contextual memory, discussed in section (4.1.1), to offer personalized assistance by dynamically assessing user preferences, behavior, and task progression. We envision these tools to adjust the level of AI support in real time, continuously aligning assistance with the user's evolving needs to enhance both efficiency and usability. These adjustments should be transparent, with the AI actively communicating its actions to the user. To preserve user autonomy, the system should provide override mechanisms, allowing users to manually adjust the level of AI assistance through simple, intuitive controls (such as sliders and buttons) ensuring a balance between AI assistance and user control. Future research could focus on determining the optimal number and type of inputs required by an AI system to ensure users retain agency and a sense of control.

\subsubsection{Future of Multimodal Prompting}
Our findings suggest that multimodal prompting significantly reduced the intentionality gap~\cite{subramonyam2024bridging} in our participants, who were primarily, novice designers. This approach enabled users to convey their ideas more accurately to AI systems, as noted in prior research~\cite{PromptCharm}.

Looking ahead, we see two key advancements in multimodality:
\begin{itemize}
\item {Towards a more natural conversation with AI:} In the near term, users will be able to provide multimodal input to AI systems through widget-based interfaces \cite{10.1145/3586183.3606777}. This process involves decomposing multifaceted user intentions or complex design requirements into discrete, manageable components via a guided interface-offering visual affordances. 

Looking ahead, we envision a future where interactions with AI transcend these structured input methods, incorporating the nuances of human communication. The integration of voice, hand gestures, facial expressions, and other subtle cues has the potential to further bridge the gap between user intention and effective human-AI collaboration. 

\item {Multimodal Sensing:} As generative AI becomes increasingly ubiquitous, it is essential to design systems that accommodate users with varying levels of domain and AI expertise. We propose that multimodal sensing emerges as a promising approach to address this challenge, extending our concept of fluid control beyond explicit inputs. Building on Schmidt's concept of implicit human-computer interaction~\cite{schmidt2000implicit,implicitinteractions_chi}, multimodal sensing enables AI systems to interpret users' emotional states, cognitive load, and engagement levels through modalities such as facial expression analysis, eye tracking, and voice pattern recognition. This aligns with our observations of SBOs diverse needs for support and control, potentially facilitating more nuanced, real-time adaptations in AI-assisted creative tasks. For instance, voice pattern analysis could gauge SBOs' emotional responses to AI suggestions, refining the system's understanding of brand alignment.

Implementation of such systems should prioritize transparency, clearly communicating to users how their unintentional actions are being interpreted and utilized in AI decision-making processes\cite{jacucci2014symbiotic}. Moreover, users should retain the ability to override the AI's interpretations and control the degree of AI involvement in their workflow~\cite{amershi2019guidelines}, ensuring that user agency remains paramount. While these technologies offer powerful capabilities for personalization, they also raise significant privacy and ethical concerns. The collection and processing of such data necessitate robust privacy safeguards and transparent data handling practices such as localised data storage, as discussed in section 5.3.

These considerations are crucial for maintaining user trust and ensuring the responsible development of multimodal sensing technologies. Future research should prioritize the development of robust algorithms for real-time analysis of implicit interactions across diverse user groups. Equally important is the design of intuitive interfaces that clearly and transparently communicate AI interpretations of these cues to users, thereby enhancing adaptive control in AI-driven co-creation processes.

\end{itemize}

\subsection{Generative AI and Data Management}

Participant questions around inputting proprietary business data into AI systems as well as the rights to the output of an AI model highlight the ongoing ambiguity around intellectual property and data management within GenAI tools. To protect business branding, tools like Etrupy \cite{entrupyHomeEntrupy} and Red Points \cite{redpointsPointsBrand} leverage AI to monitor brand integrity via a watermark - a similar capability could be integrated into ACAI \cite{li2024doubleiwatermarkprotectingmodel} and other AI-mediated design tools. Additional safeguards, including keyword filters and prompt prefixes, are methods to mitigate potential brand infringement, but there remains much work to be done to improve the efficacy of these methods \cite{10.1145/3644815.3644952, 10.14778/3681954.3681994}. 
For contextual memory, we recommend the following data management strategies: 

\begin{itemize}
\item [1] \textit{{Local Data Control}}: Storing data locally allows SBOs to edit and organize the information that feeds into the model, enhancing their control over the AI-assisted design process.
\item [2] \textit{{Improved Transparency}}: Localized data storage gives SBOs a clearer understanding of how data is utilized in AI decision-making, thereby increasing the system’s overall transparency.
\end{itemize}
Future research could also explore data management for multimodal sensing capabilities. Users like SBOs have limited bandwidth to learn about the trade offs of different privacy and management techniques on top of new Gen AI technologies. Additional research is needed to clarify the relationships between potential ease of use, utility, and privacy concerns for busy users open to new modalities \cite{10.1145/3593227}.

\section{Limitations}

While our study provides valuable insights into SBOs' interactions with ACAI, we acknowledge several limitations in this section to contextualize our findings and propose directions for future research. First, our study focused on SBOs in London with varying years of business ownership, industry backgrounds, and levels of AI familiarity. Future research could benefit from targeting a more homogeneous sample, such as SBOs within the same industry, to better understand how ACAI generates distinct yet brand-aligned ads for specific sectors. Second, while participants expressed trust in ACAI for data handling, we recognize that the controlled study setting and informed consent process may have positively influenced their perceptions. Future studies could examine user interactions with ACAI in more naturalistic environments to provide a deeper understanding of real-world trust and engagement patterns. Third, due to prototype constraints, participants were shown only a single visual advertisement. Future iterations could incorporate multiple ad outputs to allow for a more thorough evaluation of ACAI's capabilities. Lastly, our study focused on user interaction and usability but did not assess the real-world performance of AI-generated ads. Future research could include concrete performance metrics, such as click-through rates, views, and conversions, to measure the practical impact of AI-generated advertisements.

\section{Conclusion}

In this study, we investigated how GenAI tools can better support small business owners (SBOs) in the creation of brand-aligned advertisements. To this end, we developed ACAI, a multimodal GenAI-powered ad creation tool, and conducted empirical research to examine its interaction with SBOs. Our findings revealed that ACAI’s structured input scaffolding and multimodal prompting mechanisms significantly enhanced SBOs' sense of co-creation. These features enabled SBOs, many of whom are often novice designers, to bridge skill gaps by providing an intuitive interface that effectively leveraged their existing business assets in the ad creation process. Our work contributes to the growing discourse on AI-mediated co-creation by centering on novice users, shifting the focus from expert users. Finally, based on our findings, we propose practical design recommendations for building more inclusive and effective AI co-creation tools, emphasizing the need for capabilities such as contextual memory, adaptive interfaces, and robust data management to better align AI systems with SBO’s needs.

\section{Acknowledgements}

We would like to thank our designer, Stevie Yip, for being a most extraordinary wizard. Additional thanks to our colleagues Bea Alessio, Iris Qu, Lucy Boyd Schachter, Mark Bowers, and Meethu Malu for your support with this paper.

\bibliography{main}

\begin{thebibliography}{67}
\providecommand{\natexlab}[1]{#1}
\providecommand{\url}[1]{\texttt{#1}}
\expandafter\ifx\csname urlstyle\endcsname\relax
  \providecommand{\doi}[1]{doi: #1}\else
  \providecommand{\doi}{doi: \begingroup \urlstyle{rm}\Url}\fi

\bibitem[Alipour et~al.(2023)Alipour, Moghaddam, Vaidhyanathan, and Kj\ae{}rgaard]{10.1145/3593227}
M.~Alipour, M.~T. Moghaddam, K.~Vaidhyanathan, and M.~B. Kj\ae{}rgaard.
\newblock Emoticontrol: Emotions-based control of user-interfaces adaptations.
\newblock \emph{Proc. ACM Hum.-Comput. Interact.}, 7\penalty0 (EICS), jun 2023.
\newblock \doi{10.1145/3593227}.
\newblock URL \url{https://doi.org/10.1145/3593227}.

\bibitem[Amershi et~al.(2019)Amershi, Weld, Vorvoreanu, Fourney, Nushi, Collisson, Suh, Iqbal, Bennett, Inkpen, et~al.]{amershi2019guidelines}
S.~Amershi, D.~Weld, M.~Vorvoreanu, A.~Fourney, B.~Nushi, P.~Collisson, J.~Suh, S.~Iqbal, P.~N. Bennett, K.~Inkpen, et~al.
\newblock Guidelines for human-ai interaction.
\newblock In \emph{Proceedings of the 2019 chi conference on human factors in computing systems}, pages 1--13, 2019.

\bibitem[Anderson et~al.(2024)Anderson, Shah, and Kreminski]{CST_Homogenization_Analysis}
B.~R. Anderson, J.~H. Shah, and M.~Kreminski.
\newblock Evaluating creativity support tools via homogenization analysis.
\newblock In \emph{Extended Abstracts of the 2024 CHI Conference on Human Factors in Computing Systems}, CHI EA '24, New York, NY, USA, 2024. Association for Computing Machinery.
\newblock ISBN 9798400703317.
\newblock \doi{10.1145/3613905.3651088}.
\newblock URL \url{https://doi.org/10.1145/3613905.3651088}.

\bibitem[Azar et~al.(2023)Azar, Rowland, Piot, Guo, Calandriello, Valko, and Munos]{azar2023generaltheoreticalparadigmunderstand}
M.~G. Azar, M.~Rowland, B.~Piot, D.~Guo, D.~Calandriello, M.~Valko, and R.~Munos.
\newblock A general theoretical paradigm to understand learning from human preferences, 2023.
\newblock URL \url{https://arxiv.org/abs/2310.12036}.

\bibitem[Ballard(2016)]{paradoxmarketingHatsWorn}
J.~Ballard.
\newblock 6 {H}ats {W}orn {B}y {B}usiness {O}wners --- paradoxmarketing.io.
\newblock \url{https://paradoxmarketing.io/capabilities/knowledge-management/insights/6-hats-worn-by-business-owners/}, 2016.
\newblock [Accessed 02-09-2024].

\bibitem[Berthon et~al.(2008)Berthon, Ewing, and Napoli]{berthon2008brand}
P.~Berthon, M.~T. Ewing, and J.~Napoli.
\newblock Brand management in small to medium-sized enterprises.
\newblock \emph{Journal of small business management}, 46\penalty0 (1):\penalty0 27--45, 2008.

\bibitem[Boden(2004)]{traditionaldef_creativity2}
M.~A. Boden.
\newblock The creative mind: Myths and mechanisms, 2004.

\bibitem[Chakrabarty et~al.(2024)Chakrabarty, Laban, Agarwal, Muresan, and Wu]{Art_or_artiface_falsePromise}
T.~Chakrabarty, P.~Laban, D.~Agarwal, S.~Muresan, and C.-S. Wu.
\newblock Art or artifice? large language models and the false promise of creativity.
\newblock In \emph{Proceedings of the CHI Conference on Human Factors in Computing Systems}, CHI '24, New York, NY, USA, 2024. Association for Computing Machinery.
\newblock ISBN 9798400703300.
\newblock \doi{10.1145/3613904.3642731}.
\newblock URL \url{https://doi.org/10.1145/3613904.3642731}.

\bibitem[Chung and Adar(2023)]{10.1145/3586183.3606777}
J.~J.~Y. Chung and E.~Adar.
\newblock Promptpaint: Steering text-to-image generation through paint medium-like interactions.
\newblock In \emph{Proceedings of the 36th Annual ACM Symposium on User Interface Software and Technology}, UIST '23, New York, NY, USA, 2023. Association for Computing Machinery.
\newblock ISBN 9798400701320.
\newblock \doi{10.1145/3586183.3606777}.
\newblock URL \url{https://doi.org/10.1145/3586183.3606777}.

\bibitem[Dahlb\"{a}ck et~al.(1993)Dahlb\"{a}ck, J\"{o}nsson, and Ahrenberg]{10.1145/169891.169968}
N.~Dahlb\"{a}ck, A.~J\"{o}nsson, and L.~Ahrenberg.
\newblock Wizard of oz studies: why and how.
\newblock In \emph{Proceedings of the 1st International Conference on Intelligent User Interfaces}, IUI '93, page 193–200, New York, NY, USA, 1993. Association for Computing Machinery.
\newblock ISBN 0897915569.
\newblock \doi{10.1145/169891.169968}.
\newblock URL \url{https://doi.org/10.1145/169891.169968}.

\bibitem[Draxler et~al.(2024)Draxler, Werner, Lehmann, Hoppe, Schmidt, Buschek, and Welsch]{ghostwriterai}
F.~Draxler, A.~Werner, F.~Lehmann, M.~Hoppe, A.~Schmidt, D.~Buschek, and R.~Welsch.
\newblock The ai ghostwriter effect: When users do not perceive ownership of ai-generated text but self-declare as authors.
\newblock \emph{ACM Transactions on Computer-Human Interaction}, 31\penalty0 (2):\penalty0 1--40, 2024.

\bibitem[Entrupy(2024)]{entrupyHomeEntrupy}
Entrupy.
\newblock {H}ome - {E}ntrupy --- entrupy.com.
\newblock \url{https://www.entrupy.com/}, 2024.
\newblock [Accessed 12-09-2024].

\bibitem[FSB(2024)]{fsbRedefiningIntelligence}
T.~F. o. S.~B. FSB.
\newblock {R}edefining {I}ntelligence --- fsb.org.uk.
\newblock \url{https://www.fsb.org.uk/resource-report/redefining-intelligence.html}, 2024.
\newblock [Accessed 30-08-2024].

\bibitem[Glaser and Strauss(2017)]{glaser2017discovery}
B.~Glaser and A.~Strauss.
\newblock \emph{Discovery of grounded theory: Strategies for qualitative research}.
\newblock Routledge, 2017.

\bibitem[Gov.uk()]{undefinedWriteBusiness}
Gov.uk.
\newblock {W}rite a business plan --- gov.uk.
\newblock \url{https://www.gov.uk/write-business-plan}.
\newblock [Accessed 02-09-2024].

\bibitem[Hartmann et~al.(2024)Hartmann, Exner, and Domdey]{hartmann2024power}
J.~Hartmann, Y.~Exner, and S.~Domdey.
\newblock The power of generative marketing: Can generative ai create superhuman visual marketing content?
\newblock 2024.

\bibitem[Hou et~al.(2024)Hou, Tamoto, and Miyashita]{myagentunderstandmebetter}
Y.~Hou, H.~Tamoto, and H.~Miyashita.
\newblock "my agent understands me better": Integrating dynamic human-like memory recall and consolidation in llm-based agents.
\newblock In \emph{Extended Abstracts of the 2024 CHI Conference on Human Factors in Computing Systems}, CHI EA '24, New York, NY, USA, 2024. Association for Computing Machinery.
\newblock ISBN 9798400703317.
\newblock \doi{10.1145/3613905.3650839}.
\newblock URL \url{https://doi.org/10.1145/3613905.3650839}.

\bibitem[Hwang(2022)]{tooLatetobecreative}
A.~H.-C. Hwang.
\newblock Too late to be creative? ai-empowered tools in creative processes.
\newblock In \emph{Extended Abstracts of the 2022 CHI Conference on Human Factors in Computing Systems}, CHI EA '22, New York, NY, USA, 2022. Association for Computing Machinery.
\newblock ISBN 9781450391566.
\newblock \doi{10.1145/3491101.3503549}.
\newblock URL \url{https://doi.org/10.1145/3491101.3503549}.

\bibitem[Inie et~al.(2023)Inie, Falk, and Tanimoto]{Designing_Participatory_AI}
N.~Inie, J.~Falk, and S.~Tanimoto.
\newblock Designing participatory ai: Creative professionals’ worries and expectations about generative ai.
\newblock In \emph{Extended Abstracts of the 2023 CHI Conference on Human Factors in Computing Systems}, CHI EA '23, New York, NY, USA, 2023. Association for Computing Machinery.
\newblock ISBN 9781450394222.
\newblock \doi{10.1145/3544549.3585657}.
\newblock URL \url{https://doi.org/10.1145/3544549.3585657}.

\bibitem[Jacucci et~al.(2014)Jacucci, Spagnolli, Freeman, and Gamberini]{jacucci2014symbiotic}
G.~Jacucci, A.~Spagnolli, J.~Freeman, and L.~Gamberini.
\newblock Symbiotic interaction: a critical definition and comparison to other human-computer paradigms.
\newblock In \emph{Symbiotic Interaction: Third International Workshop, Symbiotic 2014, Helsinki, Finland, October 30-31, 2014, Proceedings 3}, pages 3--20. Springer, 2014.

\bibitem[Jiang et~al.(2023)Jiang, Brown, Cheng, Khan, Gupta, Workman, Hanna, Flowers, and Gebru]{AI_art_and_impact}
H.~H. Jiang, L.~Brown, J.~Cheng, M.~Khan, A.~Gupta, D.~Workman, A.~Hanna, J.~Flowers, and T.~Gebru.
\newblock Ai art and its impact on artists.
\newblock In \emph{Proceedings of the 2023 AAAI/ACM Conference on AI, Ethics, and Society}, AIES '23, page 363–374, New York, NY, USA, 2023. Association for Computing Machinery.
\newblock ISBN 9798400702310.
\newblock \doi{10.1145/3600211.3604681}.
\newblock URL \url{https://doi.org/10.1145/3600211.3604681}.

\bibitem[Kobiella et~al.(2024{\natexlab{a}})Kobiella, Flores~L\'{o}pez, Waltenberger, Draxler, and Schmidt]{If_the_Machine_Is_As_Good_As_Me_Then_What_Use_Am_I?}
C.~Kobiella, Y.~S. Flores~L\'{o}pez, F.~Waltenberger, F.~Draxler, and A.~Schmidt.
\newblock "if the machine is as good as me, then what use am i?" – how the use of chatgpt changes young professionals' perception of productivity and accomplishment.
\newblock In \emph{Proceedings of the CHI Conference on Human Factors in Computing Systems}, CHI '24, New York, NY, USA, 2024{\natexlab{a}}. Association for Computing Machinery.
\newblock ISBN 9798400703300.
\newblock \doi{10.1145/3613904.3641964}.
\newblock URL \url{https://doi.org/10.1145/3613904.3641964}.

\bibitem[Kobiella et~al.(2024{\natexlab{b}})Kobiella, Flores~L\'{o}pez, Waltenberger, Draxler, and Schmidt]{postprocessing}
C.~Kobiella, Y.~S. Flores~L\'{o}pez, F.~Waltenberger, F.~Draxler, and A.~Schmidt.
\newblock "if the machine is as good as me, then what use am i?" – how the use of chatgpt changes young professionals' perception of productivity and accomplishment.
\newblock In \emph{Proceedings of the CHI Conference on Human Factors in Computing Systems}, CHI '24, New York, NY, USA, 2024{\natexlab{b}}. Association for Computing Machinery.
\newblock ISBN 9798400703300.
\newblock \doi{10.1145/3613904.3641964}.
\newblock URL \url{https://doi.org/10.1145/3613904.3641964}.

\bibitem[Kotturi et~al.(2022)Kotturi, Johnson, Skirpan, Fox, Bigham, and Pavel]{10.1145/3491102.3517708}
Y.~Kotturi, H.~T. Johnson, M.~Skirpan, S.~E. Fox, J.~P. Bigham, and A.~Pavel.
\newblock Tech help desk: Support for local entrepreneurs addressing the long tail of computing challenges.
\newblock In \emph{Proceedings of the 2022 CHI Conference on Human Factors in Computing Systems}, CHI '22, New York, NY, USA, 2022. Association for Computing Machinery.
\newblock ISBN 9781450391573.
\newblock \doi{10.1145/3491102.3517708}.
\newblock URL \url{https://doi.org/10.1145/3491102.3517708}.

\bibitem[Kotturi et~al.(2024)Kotturi, Anderson, Ford, Skirpan, and Bigham]{VeneerofSimplicity_Entrepreneurs}
Y.~Kotturi, A.~Anderson, G.~Ford, M.~Skirpan, and J.~P. Bigham.
\newblock Deconstructing the veneer of simplicity: Co-designing introductory generative ai workshops with local entrepreneurs.
\newblock In \emph{Proceedings of the CHI Conference on Human Factors in Computing Systems}, CHI '24, New York, NY, USA, 2024. Association for Computing Machinery.
\newblock ISBN 9798400703300.
\newblock \doi{10.1145/3613904.3642191}.
\newblock URL \url{https://doi.org/10.1145/3613904.3642191}.

\bibitem[Kraljic and Lahav(2024)]{fromPrompt_toCollab}
T.~Kraljic and M.~Lahav.
\newblock From prompt engineering to collaborating: A human-centered approach to ai interfaces.
\newblock \emph{Interactions}, 31\penalty0 (3):\penalty0 30–35, may 2024.
\newblock ISSN 1072-5520.
\newblock \doi{10.1145/3652622}.
\newblock URL \url{https://doi.org/10.1145/3652622}.

\bibitem[Kutz(2008)]{contextualintelligence}
M.~Kutz.
\newblock Toward a conceptual model of contextual intelligence: A transferable leadership construct.
\newblock \emph{Kravis Leadership Institute Leadership Review}, 8:\penalty0 18--31, 01 2008.

\bibitem[Lee et~al.(2024)Lee, Cooper, and Grimmelmann]{10.1145/3614407.3643696}
K.~Lee, A.~F. Cooper, and J.~Grimmelmann.
\newblock Talkin' 'bout ai generation: Copyright and the generative-ai supply chain (the short version).
\newblock In \emph{Proceedings of the Symposium on Computer Science and Law}, CSLAW '24, page 48–63, New York, NY, USA, 2024. Association for Computing Machinery.
\newblock ISBN 9798400703331.
\newblock \doi{10.1145/3614407.3643696}.
\newblock URL \url{https://doi.org/10.1145/3614407.3643696}.

\bibitem[Li et~al.(2024{\natexlab{a}})Li, Hong, Xie, Tan, Xin, Hou, Yin, Wang, Hendrycks, Wang, Li, He, and Song]{10.14778/3681954.3681994}
Q.~Li, J.~Hong, C.~Xie, J.~Tan, R.~Xin, J.~Hou, X.~Yin, Z.~Wang, D.~Hendrycks, Z.~Wang, B.~Li, B.~He, and D.~Song.
\newblock Llm-pbe: Assessing data privacy in large language models.
\newblock \emph{Proc. VLDB Endow.}, 17\penalty0 (11):\penalty0 3201–3214, aug 2024{\natexlab{a}}.
\newblock ISSN 2150-8097.
\newblock \doi{10.14778/3681954.3681994}.
\newblock URL \url{https://doi.org/10.14778/3681954.3681994}.

\bibitem[Li et~al.(2024{\natexlab{b}})Li, Yao, Gao, Zhang, and Li]{li2024doubleiwatermarkprotectingmodel}
S.~Li, L.~Yao, J.~Gao, L.~Zhang, and Y.~Li.
\newblock Double-i watermark: Protecting model copyright for llm fine-tuning, 2024{\natexlab{b}}.
\newblock URL \url{https://arxiv.org/abs/2402.14883}.

\bibitem[Li et~al.(2024{\natexlab{c}})Li, Liang, Peng, and Yin]{valueBenefitConcernOfGenAI}
Z.~Li, C.~Liang, J.~Peng, and M.~Yin.
\newblock The value, benefits, and concerns of generative ai-powered assistance in writing.
\newblock In \emph{Proceedings of the CHI Conference on Human Factors in Computing Systems}, CHI '24, New York, NY, USA, 2024{\natexlab{c}}. Association for Computing Machinery.
\newblock ISBN 9798400703300.
\newblock \doi{10.1145/3613904.3642625}.
\newblock URL \url{https://doi.org/10.1145/3613904.3642625}.

\bibitem[Lim et~al.(2024)Lim, Cho, Kim, Park, Shin, Choi, Park, Lee, Kim, Lee, and Hong]{trust}
H.~Lim, J.~Y. Cho, T.~Kim, J.~Park, H.~Shin, S.~Choi, S.~Park, K.~Lee, J.~Kim, M.~Lee, and H.~Hong.
\newblock Co-creating question-and-answer style articles with large language models for research promotion.
\newblock In \emph{Proceedings of the 2024 ACM Designing Interactive Systems Conference}, DIS '24, page 975–994, New York, NY, USA, 2024. Association for Computing Machinery.
\newblock ISBN 9798400705830.
\newblock \doi{10.1145/3643834.3660705}.
\newblock URL \url{https://doi.org/10.1145/3643834.3660705}.

\bibitem[Liu et~al.(2024)Liu, Liu, Fiannaca, Koo, Dixon, Terry, and Cai]{10.1145/3613905.3650756}
M.~X. Liu, F.~Liu, A.~J. Fiannaca, T.~Koo, L.~Dixon, M.~Terry, and C.~J. Cai.
\newblock "we need structured output": Towards user-centered constraints on large language model output.
\newblock In \emph{Extended Abstracts of the 2024 CHI Conference on Human Factors in Computing Systems}, CHI EA '24, New York, NY, USA, 2024. Association for Computing Machinery.
\newblock ISBN 9798400703317.
\newblock \doi{10.1145/3613905.3650756}.
\newblock URL \url{https://doi.org/10.1145/3613905.3650756}.

\bibitem[Liu and Chilton(2022)]{designpromptengineering}
V.~Liu and L.~B. Chilton.
\newblock Design guidelines for prompt engineering text-to-image generative models.
\newblock In \emph{Proceedings of the 2022 CHI Conference on Human Factors in Computing Systems}, CHI '22, New York, NY, USA, 2022. Association for Computing Machinery.
\newblock ISBN 9781450391573.
\newblock \doi{10.1145/3491102.3501825}.
\newblock URL \url{https://doi.org/10.1145/3491102.3501825}.

\bibitem[Louie et~al.(2020)Louie, Coenen, Huang, Terry, and Cai]{CaiCoCoCo}
R.~Louie, A.~Coenen, C.~Z. Huang, M.~Terry, and C.~J. Cai.
\newblock Novice-ai music co-creation via ai-steering tools for deep generative models.
\newblock In \emph{Proceedings of the 2020 CHI Conference on Human Factors in Computing Systems}, CHI '20, page 1–13, New York, NY, USA, 2020. Association for Computing Machinery.
\newblock ISBN 9781450367080.
\newblock \doi{10.1145/3313831.3376739}.
\newblock URL \url{https://doi.org/10.1145/3313831.3376739}.

\bibitem[Lyu et~al.(2024)Lyu, Zhang, Niu, and Cai]{Youtubers_Use_Of_GenAI}
Y.~Lyu, H.~Zhang, S.~Niu, and J.~Cai.
\newblock A preliminary exploration of youtubers' use of generative-ai in content creation.
\newblock In \emph{Extended Abstracts of the 2024 CHI Conference on Human Factors in Computing Systems}, CHI EA '24, New York, NY, USA, 2024. Association for Computing Machinery.
\newblock ISBN 9798400703317.
\newblock \doi{10.1145/3613905.3651057}.
\newblock URL \url{https://doi.org/10.1145/3613905.3651057}.

\bibitem[Ma et~al.(2024)Ma, Mishra, Liu, Su, Chen, Kulkarni, Cheng, Le, and Chi]{10.1145/3613905.3651093}
X.~Ma, S.~Mishra, A.~Liu, S.~Y. Su, J.~Chen, C.~Kulkarni, H.-T. Cheng, Q.~Le, and E.~Chi.
\newblock Beyond chatbots: Explorellm for structured thoughts and personalized model responses.
\newblock In \emph{Extended Abstracts of the 2024 CHI Conference on Human Factors in Computing Systems}, CHI EA '24, New York, NY, USA, 2024. Association for Computing Machinery.
\newblock ISBN 9798400703317.
\newblock \doi{10.1145/3613905.3651093}.
\newblock URL \url{https://doi.org/10.1145/3613905.3651093}.

\bibitem[Mahdavi~Goloujeh et~al.(2024)Mahdavi~Goloujeh, Sullivan, and Magerko]{isITai_orMe}
A.~Mahdavi~Goloujeh, A.~Sullivan, and B.~Magerko.
\newblock Is it ai or is it me? understanding users’ prompt journey with text-to-image generative ai tools.
\newblock In \emph{Proceedings of the CHI Conference on Human Factors in Computing Systems}, CHI '24, New York, NY, USA, 2024. Association for Computing Machinery.
\newblock ISBN 9798400703300.
\newblock \doi{10.1145/3613904.3642861}.
\newblock URL \url{https://doi.org/10.1145/3613904.3642861}.

\bibitem[Mirowski et~al.(2023)Mirowski, Mathewson, Pittman, and Evans]{dramatron_deepmind}
P.~Mirowski, K.~W. Mathewson, J.~Pittman, and R.~Evans.
\newblock Co-writing screenplays and theatre scripts with language models: Evaluation by industry professionals.
\newblock In \emph{Proceedings of the 2023 CHI Conference on Human Factors in Computing Systems}, pages 1--34, 2023.

\bibitem[Mirowski et~al.(2024)Mirowski, Love, Mathewson, and Mohamed]{comedy_deepmind}
P.~Mirowski, J.~Love, K.~Mathewson, and S.~Mohamed.
\newblock A robot walks into a bar: Can language models serve as creativity supporttools for comedy? an evaluation of llms’ humour alignment with comedians.
\newblock In \emph{The 2024 ACM Conference on Fairness, Accountability, and Transparency}, pages 1622--1636, 2024.

\bibitem[Oh et~al.(2024)Oh, Kim, and Kim]{lumimood}
J.~Oh, S.~Kim, and S.~Kim.
\newblock Lumimood: A creativity support tool for designing the mood of a 3d scene.
\newblock In \emph{Proceedings of the CHI Conference on Human Factors in Computing Systems}, CHI '24, New York, NY, USA, 2024. Association for Computing Machinery.
\newblock ISBN 9798400703300.
\newblock \doi{10.1145/3613904.3642440}.
\newblock URL \url{https://doi.org/10.1145/3613904.3642440}.

\bibitem[Peng et~al.(2024)Peng, Koch, and Mackay]{10.1145/3643834.3661588}
X.~Peng, J.~Koch, and W.~E. Mackay.
\newblock Designprompt: Using multimodal interaction for design exploration with generative ai.
\newblock In \emph{Proceedings of the 2024 ACM Designing Interactive Systems Conference}, DIS '24, page 804–818, New York, NY, USA, 2024. Association for Computing Machinery.
\newblock ISBN 9798400705830.
\newblock \doi{10.1145/3643834.3661588}.
\newblock URL \url{https://doi.org/10.1145/3643834.3661588}.

\bibitem[Points()]{redpointsPointsBrand}
R.~Points.
\newblock {R}ed {P}oints’ {A}{I} {B}rand {P}rotection {P}latform --- redpoints.com.
\newblock \url{https://www.redpoints.com/}, year = {2024}, note = {[Accessed 12-09-2024]},.

\bibitem[Pratt(2024)]{forbesSmallBusiness}
K.~Pratt.
\newblock {U}{K} {S}mall {B}usiness {S}tatistics --- forbes.com.
\newblock \url{https://www.forbes.com/uk/advisor/business/small-business-statistics/}, 2024.
\newblock [Accessed 30-08-2024].

\bibitem[Radlinski et~al.(2022)Radlinski, Balog, Diaz, Dixon, and Wedin]{10.1145/3477495.3531873}
F.~Radlinski, K.~Balog, F.~Diaz, L.~Dixon, and B.~Wedin.
\newblock On natural language user profiles for transparent and scrutable recommendation.
\newblock In \emph{Proceedings of the 45th International ACM SIGIR Conference on Research and Development in Information Retrieval}, SIGIR '22, page 2863–2874, New York, NY, USA, 2022. Association for Computing Machinery.
\newblock ISBN 9781450387323.
\newblock \doi{10.1145/3477495.3531873}.
\newblock URL \url{https://doi.org/10.1145/3477495.3531873}.

\bibitem[Rezwana and Maher(2022)]{identifyingethicalissuesai}
J.~Rezwana and M.~L. Maher.
\newblock Identifying ethical issues in ai partners in human-ai co-creation.
\newblock 2022.
\newblock URL \url{https://arxiv.org/abs/2204.07644}.

\bibitem[Rezwana and Maher(2023)]{designfictionstudy}
J.~Rezwana and M.~L. Maher.
\newblock User perspectives on ethical challenges in human-ai co-creativity: A design fiction study.
\newblock In \emph{Proceedings of the 15th Conference on Creativity and Cognition}, page 62–74, New York, NY, USA, 2023. Association for Computing Machinery.
\newblock ISBN 9798400701801.
\newblock \doi{10.1145/3591196.3593364}.
\newblock URL \url{https://doi.org/10.1145/3591196.3593364}.

\bibitem[Sanner et~al.(2023)Sanner, Balog, Radlinski, Wedin, and Dixon]{10.1145/3604915.3608845}
S.~Sanner, K.~Balog, F.~Radlinski, B.~Wedin, and L.~Dixon.
\newblock Large language models are competitive near cold-start recommenders for language- and item-based preferences.
\newblock In \emph{Proceedings of the 17th ACM Conference on Recommender Systems}, RecSys '23, page 890–896, New York, NY, USA, 2023. Association for Computing Machinery.
\newblock ISBN 9798400702419.
\newblock \doi{10.1145/3604915.3608845}.
\newblock URL \url{https://doi.org/10.1145/3604915.3608845}.

\bibitem[Schmidt(2000)]{schmidt2000implicit}
A.~Schmidt.
\newblock Implicit human computer interaction through context.
\newblock \emph{Personal technologies}, 4:\penalty0 191--199, 2000.

\bibitem[Serim and Jacucci(2019)]{implicitinteractions_chi}
B.~Serim and G.~Jacucci.
\newblock Explicating "implicit interaction": An examination of the concept and challenges for research.
\newblock In \emph{Proceedings of the 2019 CHI Conference on Human Factors in Computing Systems}, CHI '19, page 1–16, New York, NY, USA, 2019. Association for Computing Machinery.
\newblock ISBN 9781450359702.
\newblock \doi{10.1145/3290605.3300647}.
\newblock URL \url{https://doi.org/10.1145/3290605.3300647}.

\bibitem[Shelby et~al.(2024)Shelby, Rismani, and Rostamzadeh]{ML_ArtistFolkTheories}
R.~Shelby, S.~Rismani, and N.~Rostamzadeh.
\newblock Generative ai in creative practice: Ml-artist folk theories of t2i use, harm, and harm-reduction.
\newblock In \emph{Proceedings of the CHI Conference on Human Factors in Computing Systems}, CHI '24, New York, NY, USA, 2024. Association for Computing Machinery.
\newblock ISBN 9798400703300.
\newblock \doi{10.1145/3613904.3642461}.
\newblock URL \url{https://doi.org/10.1145/3613904.3642461}.

\bibitem[Shi et~al.(2023)Shi, Gao, Jiao, and Cao]{Design_Collab_b/w_Designers_AI_LiteratureReview}
Y.~Shi, T.~Gao, X.~Jiao, and N.~Cao.
\newblock Understanding design collaboration between designers and artificial intelligence: A systematic literature review.
\newblock \emph{Proc. ACM Hum.-Comput. Interact.}, 7\penalty0 (CSCW2), oct 2023.
\newblock \doi{10.1145/3610217}.
\newblock URL \url{https://doi.org/10.1145/3610217}.

\bibitem[Simonton(2012)]{traditionaldef_creativity}
D.~K. Simonton.
\newblock Taking the us patent office criteria seriously: A quantitative three-criterion creativity definition and its implications.
\newblock \emph{Creativity research journal}, 24\penalty0 (2-3):\penalty0 97--106, 2012.

\bibitem[Subramonyam et~al.(2024)Subramonyam, Pea, Pondoc, Agrawala, and Seifert]{subramonyam2024bridging}
H.~Subramonyam, R.~Pea, C.~Pondoc, M.~Agrawala, and C.~Seifert.
\newblock Bridging the gulf of envisioning: Cognitive challenges in prompt based interactions with llms.
\newblock In \emph{Proceedings of the CHI Conference on Human Factors in Computing Systems}, pages 1--19, 2024.

\bibitem[Sun et~al.(2024)Sun, Jang, Ma, and Wang]{genai_wild}
Y.~Sun, E.~Jang, F.~Ma, and T.~Wang.
\newblock Generative ai in the wild: Prospects, challenges, and strategies.
\newblock In \emph{Proceedings of the CHI Conference on Human Factors in Computing Systems}, CHI '24, New York, NY, USA, 2024. Association for Computing Machinery.
\newblock ISBN 9798400703300.
\newblock \doi{10.1145/3613904.3642160}.
\newblock URL \url{https://doi.org/10.1145/3613904.3642160}.

\bibitem[Torricelli et~al.(2024)Torricelli, Martino, Baronchelli, and Aiello]{10.1145/3614419.3644000}
M.~Torricelli, M.~Martino, A.~Baronchelli, and L.~M. Aiello.
\newblock The role of interface design on prompt-mediated creativity in generative ai.
\newblock In \emph{Proceedings of the 16th ACM Web Science Conference}, WEBSCI '24, page 235–240, New York, NY, USA, 2024. Association for Computing Machinery.
\newblock ISBN 9798400703348.
\newblock \doi{10.1145/3614419.3644000}.
\newblock URL \url{https://doi.org/10.1145/3614419.3644000}.

\bibitem[Verheijden and Funk(2023)]{cocoDifussion}
M.~P. Verheijden and M.~Funk.
\newblock Collaborative diffusion: Boosting designerly co-creation with generative ai.
\newblock In \emph{Extended Abstracts of the 2023 CHI Conference on Human Factors in Computing Systems}, CHI EA '23, New York, NY, USA, 2023. Association for Computing Machinery.
\newblock ISBN 9781450394222.
\newblock \doi{10.1145/3544549.3585680}.
\newblock URL \url{https://doi.org/10.1145/3544549.3585680}.

\bibitem[Vygotsky(1978)]{vygotsky1978mind}
L.~S. Vygotsky.
\newblock \emph{Mind in society: The development of higher psychological processes}, volume~86.
\newblock Harvard university press, 1978.

\bibitem[Wadinambiarachchi et~al.(2024)Wadinambiarachchi, Kelly, Pareek, Zhou, and Velloso]{Divergent_Thinking}
S.~Wadinambiarachchi, R.~M. Kelly, S.~Pareek, Q.~Zhou, and E.~Velloso.
\newblock The effects of generative ai on design fixation and divergent thinking.
\newblock In \emph{Proceedings of the CHI Conference on Human Factors in Computing Systems}, CHI '24, New York, NY, USA, 2024. Association for Computing Machinery.
\newblock ISBN 9798400703300.
\newblock \doi{10.1145/3613904.3642919}.
\newblock URL \url{https://doi.org/10.1145/3613904.3642919}.

\bibitem[Wang et~al.(2024)Wang, Huang, Song, Ma, and Zhang]{PromptCharm}
Z.~Wang, Y.~Huang, D.~Song, L.~Ma, and T.~Zhang.
\newblock Promptcharm: Text-to-image generation through multi-modal prompting and refinement.
\newblock In \emph{Proceedings of the CHI Conference on Human Factors in Computing Systems}, CHI '24, New York, NY, USA, 2024. Association for Computing Machinery.
\newblock ISBN 9798400703300.
\newblock \doi{10.1145/3613904.3642803}.
\newblock URL \url{https://doi.org/10.1145/3613904.3642803}.

\bibitem[Welsh and White()]{hbrSmallBusiness}
J.~A. Welsh and J.~F. White.
\newblock {A} {S}mall {B}usiness {I}s {N}ot a {L}ittle {B}ig {B}usiness --- hbr.org.
\newblock \url{https://hbr.org/1981/07/a-small-business-is-not-a-little-big-business}.
\newblock [Accessed 02-09-2024].

\bibitem[Wu et~al.(2021)Wu, Ji, Yu, Zeng, Wu, and Shidujaman]{10.1007/978-3-030-78462-1_13}
Z.~Wu, D.~Ji, K.~Yu, X.~Zeng, D.~Wu, and M.~Shidujaman.
\newblock Ai creativity and the human-ai co-creation model.
\newblock In \emph{Human-Computer Interaction. Theory, Methods and Tools: Thematic Area, HCI 2021, Held as Part of the 23rd HCI International Conference, HCII 2021, Virtual Event, July 24–29, 2021, Proceedings, Part I}, page 171–190, Berlin, Heidelberg, 2021. Springer-Verlag.
\newblock ISBN 978-3-030-78461-4.
\newblock \doi{10.1007/978-3-030-78462-1_13}.
\newblock URL \url{https://doi.org/10.1007/978-3-030-78462-1_13}.

\bibitem[Yang et~al.(2023)Yang, Ongpin, Nikolenko, Huang, and Farseev]{opacity}
Q.~Yang, M.~Ongpin, S.~Nikolenko, A.~Huang, and A.~Farseev.
\newblock Against opacity: Explainable ai and large language models for effective digital advertising.
\newblock In \emph{Proceedings of the 31st ACM International Conference on Multimedia}, MM '23, page 9299–9305, New York, NY, USA, 2023. Association for Computing Machinery.
\newblock ISBN 9798400701085.
\newblock \doi{10.1145/3581783.3612817}.
\newblock URL \url{https://doi.org/10.1145/3581783.3612817}.

\bibitem[Zamfirescu-Pereira et~al.(2023)Zamfirescu-Pereira, Wong, Hartmann, and Yang]{johnnycantprompt_nonAIexperts}
J.~Zamfirescu-Pereira, R.~Y. Wong, B.~Hartmann, and Q.~Yang.
\newblock Why johnny can’t prompt: how non-ai experts try (and fail) to design llm prompts.
\newblock In \emph{Proceedings of the 2023 CHI Conference on Human Factors in Computing Systems}, pages 1--21, 2023.

\bibitem[Zhang et~al.(2024)Zhang, Xia, Liu, Xu, Hoang, Xing, Staples, Lu, and Zhu]{10.1145/3644815.3644952}
D.~Zhang, B.~Xia, Y.~Liu, X.~Xu, T.~Hoang, Z.~Xing, M.~Staples, Q.~Lu, and L.~Zhu.
\newblock Privacy and copyright protection in generative ai: A lifecycle perspective.
\newblock In \emph{Proceedings of the IEEE/ACM 3rd International Conference on AI Engineering - Software Engineering for AI}, CAIN '24, page 92–97, New York, NY, USA, 2024. Association for Computing Machinery.
\newblock ISBN 9798400705915.
\newblock \doi{10.1145/3644815.3644952}.
\newblock URL \url{https://doi.org/10.1145/3644815.3644952}.

\bibitem[Zhong et~al.(2024)Zhong, Guo, Gao, Ye, and Wang]{zhong2024memorybank}
W.~Zhong, L.~Guo, Q.~Gao, H.~Ye, and Y.~Wang.
\newblock Memorybank: Enhancing large language models with long-term memory.
\newblock In \emph{Proceedings of the AAAI Conference on Artificial Intelligence}, volume~38, pages 19724--19731, 2024.

\bibitem[Zhou et~al.(2024)Zhou, Li, Tang, Tang, Li, Cui, and Wu]{NonLinearCollaboration}
J.~Zhou, R.~Li, J.~Tang, T.~Tang, H.~Li, W.~Cui, and Y.~Wu.
\newblock Understanding nonlinear collaboration between human and ai agents: A co-design framework for creative design.
\newblock In \emph{Proceedings of the CHI Conference on Human Factors in Computing Systems}, CHI '24, New York, NY, USA, 2024. Association for Computing Machinery.
\newblock ISBN 9798400703300.
\newblock \doi{10.1145/3613904.3642812}.
\newblock URL \url{https://doi.org/10.1145/3613904.3642812}.

\end{thebibliography}

\newpage
\section{Appendix}
 \begin{figure*}[h!]
  \includegraphics[width=\linewidth]{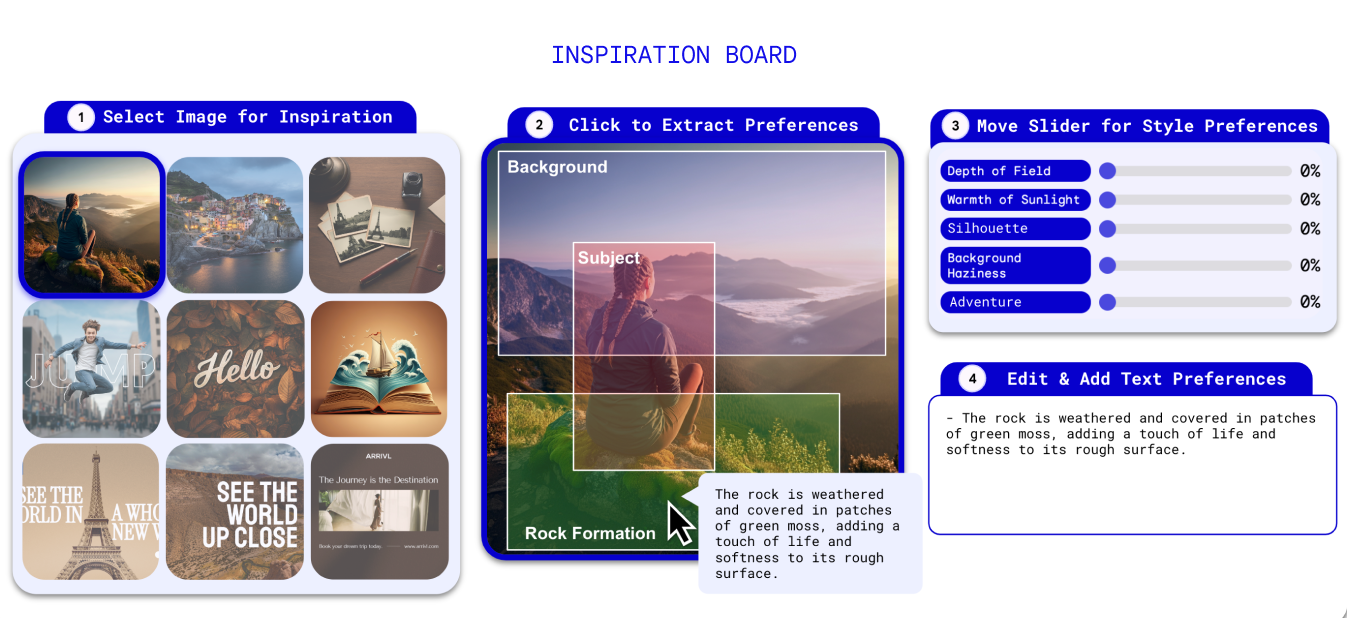}
  \caption{Detailed view of ACAI inspiration panel interface: 1) Image gallery for inspiration selection comprised of existing business images and additional stylistic images 2) Extraction of style preferences to add to super prompt 3) Sliders for stylistic preferences to add to super prompt 4) Text box to refine stylistic preferences}

 \end{figure*}

 \begin{figure*}[h!]
  \includegraphics[width=0.8\linewidth]{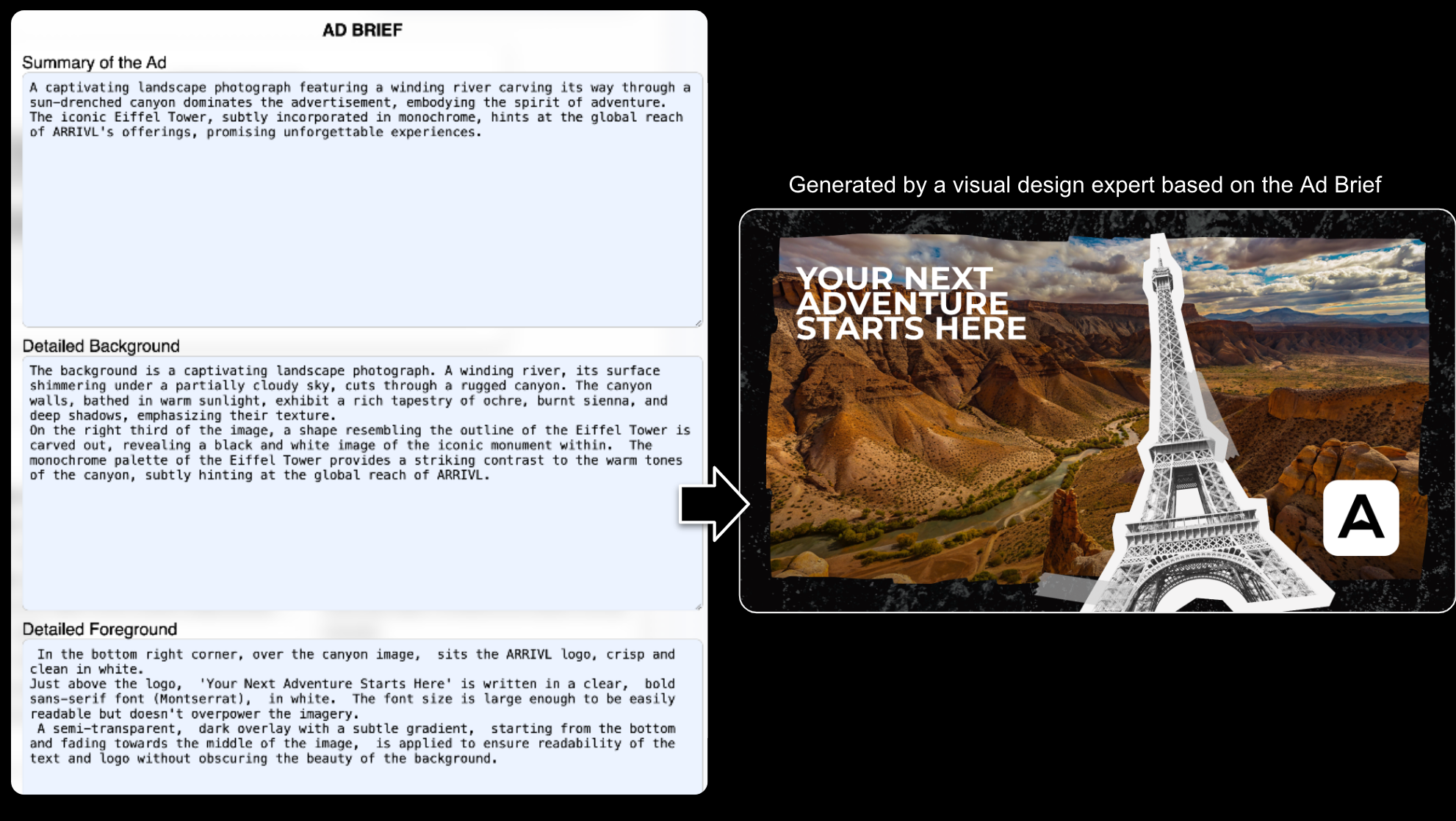}
  \caption{Detailed view of ACAI `Ad brief’ for demo business consisting of summary, detailed background, and detailed foreground to provide details to designer to wizard-of-oz visual advertisement.}

 \end{figure*}

\end{document}